\documentclass{emulateapj}

\newcommand{\be}{\begin{equation}}
\newcommand{\ee}{\end{equation}}

\newcommand{\kms}{km s$^{-1}$}

\newcommand{\ax}{$\alpha_{\rm X}$}

\newcommand{\msun}{$M_{\odot}$}

\newcommand{\xmm}{{\it XMM-Newton}}
\newcommand{\chandra}{{\it Chandra}}

\newcommand{\mJybeam}{mJy\thinspace beam$^{-1}$}
\newcommand{\atc}{atoms\thinspace cm$^{-2}$} 
\newcommand{\sixcm}{$\lambda 6$ cm}
\newcommand{\CO}{$^{12}$CO $J=1\rightarrow 0$}
\newcommand{\HI}{H {\ninerm I}}
\newcommand{\HII}{H {\ninerm II}}
\font\ninerm=cmr9
\shorttitle{Observations of NGC 2207/IC 2163}
\shortauthors{Kaufman et al.}

\begin{document}

%\input DGrupe_clipfig.tex
%\useunitmm

\def\etal{{\it et\thinspace al.}\ }
\def\alp{{$\alpha$}\ }
\def\al2{{$\alpha^2$}\ }

%\def \charthoffset {\hspace{0.2cm}} \def \charthsep {\hspace{0.3cm}}
%\def \chartvsepcap {\vspace{0.3cm}}
%\def \chartvsep {\vspace{0.1cm}}
%\newcommand{\putchartb}[1]{\clipfig{#1}{75}{20}{7}{275}{192}}
%\newcommand{\putchartc}[1]{\clipfig{#1}{55}{33}{19}{275}{195}}
%
%
%
%\newcommand{\chartlineb}[2]{\parbox[t]{18cm}{\noindent\charthoffset\putchartb{#1}\charthsep\putchartb{#2}\chartvsep}}
%\newcommand{\chartlinec}[3]{\parbox[t]{18cm}{\noindent\charthoffset\putchartc{#1}\charthsep\putchartc{#2}\chartvsep\putchartc{#3}\chartvsep}}


\title{NGC 2207/IC 2163: A Grazing Encounter with Large Scale Shocks
}


\author{Michele Kaufman\altaffilmark{1},
Dirk Grupe\altaffilmark{2},
%\email{grupe@astro.psu.edu},
Debra M. Elmegreen\altaffilmark{3},
Bruce G. Elmegreen\altaffilmark{4},
Curtis Struck\altaffilmark{5},
Elias Brinks\altaffilmark{6}
}

\altaffiltext{1}{Department of Physics, The Ohio State University, 191 West
Woodruff Ave,
Columbus, OH 43210; email: rallis@mps.ohio-state.edu}

\altaffiltext{2}{Department of Astronomy and Astrophysics, Pennsylvania State
University, 525 Davey Lab, University Park, PA 16802; email: 
grupe@astro.psu.edu} 

\altaffiltext{3}{Department of Physics and Astronomy, Vassar College, 124
Raymond Av., Poughkeepsie, NY 12604; email: elmegreen@vassar.edu
}

\altaffiltext{4}{IBM Research Division, T.J. Watson Research Center, PO Box 218,
Yorktown Heights, NY 10598; email: bge@us.ibm.com
}

\altaffiltext{5}{Department of Physics and Astronomy, Iowa State University,
Ames, IA 50011; email: curt@iastate.edu}

\altaffiltext{6}{University of Hertfordshire, Centre for Astrophysics Research,
College Lane, Hatfield AL10 9AB, UK; email: E.Brinks@herts.ac.uk}


%\altaffiltext{1}{Visiting Astronomer, Cerro Tololo Inter-American Observat}


\begin{abstract}
Radio continuum, {\it Spitzer} infrared, optical, and \xmm\ X-ray and 
ultraviolet observations ($UVW1$ and $UVM2$) are used to study 
large--scale shock fronts, 
young star complexes, and the galactic nuclei in the interacting galaxies 
NGC~2207/IC 2163. There are
two types of large--scale shock fronts in this galaxy pair. 
The large-scale shock front along the rim of the ocular oval in
IC 2163 has produced vigorous 
star formation in a dusty environment, bright in the 
{\it Spitzer} 8 \micron\ and 24 \micron\ images. This large-scale shock 
lies behind NGC 2207 and is not prominent in X-rays. 
In the outer part of the companion side of NGC 2207,   
a large-scale front attributed to disk or halo scraping is 
particularly bright in the \sixcm\ and $\lambda 20$ cm radio continuum but
not in any tracers of recent star formation (H$\alpha$, 8 \micron,
24  \micron, or ultraviolet emission) or in X-rays.  
This radio-continuum front is simply due to 
compression of the magnetic field and 
may be mainly in the halo on the back side of NGC 2207, between the two
galaxies. Values of the ratio of
8 \micron\ to \sixcm\ radio continuum flux density of
prominent, kpc-sized, {\it Spitzer} IRAC 
star-forming clumps in NGC 2207/IC 2163 are compared with those of
giant radio \HII\ regions in M81. For the bright clumps in NGC 2207, the
mean value of this ratio is the same as for the M81 \HII\ regions, whereas
for the bright clumps on the rim of the IC 2163 ocular oval, the
mean value is nearly a factor of two greater. 
On the other hand, the galaxy pair has global values of the ratios of infrared 
to radio continuum flux density in the
 {\it Spitzer} 8 \micron, 24 \micron, and 70 \micron\ bands and
the {\it IRAS} FIR  
significantly below the medians/means for large samples of galaxies.  
{\it Feature i}, a mini-starburst
on an outer arm of NGC 2207 on its anti-companion side, is the most
luminous 8 $\mu$m, 24 $\mu$m, 70 $\mu$m, radio continuum, and H$\alpha$ 
source in the galaxy pair. Highly absorbed, it
is not detected in X-rays and is fainter in the ultraviolet than the clump
containing SN 1999ec, 8\arcsec\ SSE of the core of {\it feature i}.
We find evidence that a radio supernova was present in the core of 
{\it feature i} in 2001.   X-ray
emission is detected from the nucleus of
NGC 2207 and from nine discrete sources whose X-ray luminosities
 make them possible candidates for ULXs. One of these corresponds with 
the Type Ib SN 1999ec, and another may be a radio supernova or a background 
quasar. The X-ray luminosity of the NGC 2207 nucleus is 
log $L_{\rm 0.3-10.0 keV}$ = 
40.6 [ergs s$^{-1}]$, which, together with its X-ray spectrum,
suggests that this is a highly absorbed, low-luminosity, 
Active Galactic Nucleus. 
\end{abstract}


\section{Introduction}

The spiral galaxies NGC 2207 and IC 2163 at a distance of 35 Mpc ($1''$ =170
pc) \citep{elmegreen95a} 
and partially overlapping in
projection are involved in a nearly--grazing encounter with closest approach
200 -- 400 Myr ago.
We have studied this pair 
extensively \citep{elmegreen95a,elmegreen95b, elmegreen98, elmegreen00, 
elmegreen01, elmegreen06, thomasson04, struck05} 
throughout the electromagnetic spectrum with {\it HST} WFPC2 observations 
in $U B V I$ bands, 
ground-based H$\alpha$, {\it Spitzer} IRAC (3.6 -- 8 $\mu$m) and 
MIPS (24 -- 160 $\mu$m) observations,
VLA \HI\ and radio continuum, and \CO\
(SEST) observations  and 
reproduced many of the observed features with N--body and SPH encounter 
simulations.  
Relative to IC 2163, the encounter is prograde and nearly 
in--plane, producing the observed eye-shaped (ocular) oval 
and two long tidal arms in IC 2163. Relative to NGC 2207, the encounter is 
retrograde with IC 2163 moving behind NGC 2207 towards the east. 
The
short--lived ocular phase and other features of this system set strict 
constraints on the numerical model for the encounter. Along the rim of
the ocular oval there is a large-scale shock front caused by the inflow
of gas responding to tidal torques. This observed shock is a signature of the
early stages of prograde grazing encounters. The models in 
\citet{struck05} predict that disk or halo scraping between the 
companion sides of the two galaxies
would push shocks at a few hundred km s$^{-1}$ into each other 
across a front $30''$ to $60''$ in length, with a mass transfer 
stream from IC 2163 impinging on  NGC 2207. Evidence for this 
is seen in the radio continuum image
 as enhanced radio emission 
from the outer part of the companion sides of NGC 2207 and IC 2163. 
According to the models in 
\citet{struck05}, the two galaxies
in this system will eventually merge.

To extend our previous studies of this galaxy pair, 
we observed 
NGC 2207/IC 2163 in X-rays and the ultraviolet with \xmm\ and made  
new radio continuum observations with the VLA\footnote{The National Radio 
Astronomy Observatory is a facility of the National Science Foundation 
operated under
cooperative agreement by Associated Universities, Inc. The observations are
from VLA Programs AK 368 and AK 526.} 
 at $\lambda 6$ cm 
at a resolution of $2.5\arcsec$, comparable to that of the ultraviolet and 
Spitzer IRAC
images. The goals of the X-ray observations were  (1) to 
detect the predicted soft X-ray emission from diffuse hot plasma 
at the large--scale shock fronts 
produced by the grazing encounter and (2) to determine the number,
location, and nature of bright, discrete X-ray sources in this pair. 
The \xmm\ observations presented here are the first deep X-ray observations of
NGC 2207/IC 2163. Note that this galaxy pair was not detected in the ROSAT
All-Sky Survey \citep[RASS; ][]{voges99}. 

We use our radio continuum, {\it Spitzer} infrared, 
optical, and \xmm\ X-ray, $UVW1$ (effective $\lambda$ = 2910 \AA) and
$UVM2$ (effective $\lambda$ = 2310 \AA) 
observations to study the large--scale shock 
fronts, the young star complexes, the NGC 2207 nucleus, and various discrete
X-ray sources in these galaxies. In an outer spiral arm on the anti-companion 
side of NGC 2207, there is a 
morphologically peculiar star-forming region
\citep[called {\it feature i} by][]{elmegreen00} 
which is the most luminous  H$\alpha$, radio continuum, 8 $\mu$m,
24 $\mu$m, and 70 $\mu$m source in NGC 2207/IC 2163. At 24 \micron,
it accounts for $\simeq 12\%$ \citep{elmegreen06} of the total emission from 
the galaxy pair. {\it Feature i} contains an
opaque dust cone (400 pc in projected length) aligned nearly parallel to
the minor axis of the projection of NGC 2207 into the sky plane.
We present new results on {\it feature i} and its environs. 

With a star formation rate (SFR) deduced from 
H$\alpha$ emission, IC 2163 and NGC 2207 each have 
a SFR/$M$(HI) typical of normal spiral disks. The radio continuum flux density 
of  NGC 2207/IC 2163 is about 3 times higher than expected from the
IRAS far--infrared flux \citep{elmegreen95a}, yet neither galaxy 
contains a radio-loud AGN. The global value of the 
Helou $q_{{\rm FIR}}$ parameter (the logarithm of the ratio of 
FIR to $\lambda 20$
cm radio continuum flux density) is 1.81 for NGC 2207/IC 2163, whereas 
\citet{condon92} finds the
median value of $q_{{\rm FIR}}$ for galaxies that are not radio-loud AGNs
 is $\simeq 2.3$ with an rms scatter of  $\leq 0.2$.
Explaining why the radio continuum emission from 
the galaxy pair is enhanced 
without a commensurate effect on star formation is important for 
understanding the conditions necessary for star formation in general.
Some other galaxies with a similarly low global value
of the Helou $q_{{\rm FIR}}$ parameter
are NGC 2276  \citep{hummel95} and 
the Taffy pairs UGC 12914/15 and UGC 813/6 \citep{condon93,condon02}.
Like NGC 2207, the spiral galaxy
 NGC 2276 has enhanced radio emission
in the outer part of the disk on one side of the galaxy. NGC 2207 and NGC 2276
have about the same redshift, radio continuum flux density and $B_T$.  
\citet{rasmussen06} observed the interacting galaxy pair NGC 2276/NGC
2300 with \chandra\ and found a shock-like feature in X-rays. NGC 2276 had
already been observed in X-rays with the ROSAT High-Resolution Imager
\citep{davis97}. We shall compare NGC 2207/IC 2163 with NGC 2276.

Section 2 describes our new observations (X-ray and ultraviolet from \xmm\
and $\lambda 6$ cm radio continuum from the VLA) and the data reductions.
Section 3 presents an overview of the system. Section 4 discusses star-forming
clumps prominent in the {\it Spitzer 8} $\mu$m (IRAC 4) image and/or in the
\xmm\ ultraviolet images ($UVM2$ and $UVW1$). Values of the flux density ratio
S$_\nu$(8 $\mu$m)/S$_\nu$($\lambda 6$ cm) for these kpc-sized clumps are
compared with those of giant HII regions in M81 as an example of what is
normal for an OB association. Section 5 presents our results on the
large-scale shock fronts and comments on infrared to
radio continuum ratios.
Section 6 describes our X-ray results. Section 7 compares
NGC 2207 with NGC 2276. Section 8
is devoted to {\it feature i} and its environs. Section 9 summarizes 
our conclusions. 

For this galaxy pair, we adopt the distance of 35 Mpc as in 
\citet{elmegreen95a}, who used $H_0$ = 75 kpc s$^{-1}$ Mpc$^{-1}$. 
 
\section{\label{observe} Observations and data reduction}

\subsection{\xmm\ observations}
\xmm\ observed NGC 2207/IC 2163 on 2005 August 31  
for a total of 51.6 ks. A summary of the
observations in each of the instruments onboard \xmm\ is given in  
Table\,\ref{obs_log}. The European Photon Imaging Camera (EPIC) pn
\citep{strueder01} was operated in Extended Full-Frame mode and
the two EPIC MOS \citep{turner01}  in Full-Frame
mode. All observations with the EPICs were performed with the thin filters. 
Due to some episodes of high particle background at the beginning of the
observations, part of the pn observation had to be discarded leaving a net
observing time of 34.2 ks. The MOS data, however, were not affected by this 
high background flux and we used the entire observations.

The \xmm\  data were analyzed in the standard way using the 
XMMSAS version {\it xmmsas\_20060628\_1801-7.0.0}. 
Only single and double events ({\tt PATTERN.le.4}) and
single to quadruple events ({\tt PATTERN.le.12}) were 
selected for the pn and MOS data, respectively. Events in or next to the CCD
gaps were rejected from the analysis ({\tt FLAG.eq.0}). For our
final  X-ray image, the screened event files of the pn and
MOS data were merged with the XMMSAS task {\it merge}. 
The spectra were rebinned by {\it grppha} version 3.0.0 with 20 photons 
per bin in the pn and 15 counts per bin in the two MOS cameras. 
The redistribution matrices and the auxiliary response files were 
created by the
XMMSAS tasks {\it rmfgen} and {\it arfgen}, respectively. 
 Spectral fits to the EPIC pn and MOS spectra were performed with XSPEC
version 12.5.0ac \citep{arnaud96}. All errors are 90\% confidence unless
stated otherwise. 

Throughout the paper spectral indexes are denoted as energy spectral indexes
with flux density
$F_{\nu} \propto \nu^{-\alpha}$. Spectral index has the same type of definition
for the X-ray and for the radio observations
 
For the MOS X-ray image, the FWHM of the point-spread function (PSF) 
is $\sim 5\arcsec$ on-axis.   The 4.1\arcsec\ pixel size
for the pn X-ray camera results in somewhat poorer spatial resolution 
than for the MOS cameras (1.1\arcsec\ pixels). 
According to the \xmm\ User's
Handbook, the core of the PSF for the X-ray cameras 
varies little over the energy range 0.1 - 4 keV and is somewhat 
triangular in shape for the MOS2 camera. The half energy width
(at which 50\%\ of the total energy is encircled) is $\sim 15\arcsec$ for the 
\xmm\ X-ray images.  

Comparison of the \xmm\ X-ray, radio continuum, and 
{\it Spitzer} infrared positions of the NGC 2207 nucleus
indicates that the positional accuracy of the X-ray data is about 3\arcsec. 

We also took advantage of \xmm's multi-wavelength capacity by using
the Optical Monitor \citep[OM; ][]{mason01} performing photometry in 4 filters
($B$, $U$, $UVW1$, and $UVM2$). 
We use the $UVW1$ (effective $\lambda$ = 2910 \AA) and
$UVM2$ (effective $\lambda$ = 2310 \AA) images to study prominent star-forming
clumps by comparing the ultraviolet, radio continuum, and 8 \micron\ flux
densities and H$\alpha$ fluxes.      
The $UVM2$-band, with response to the $\lambda$ range 2000\AA\
to 2700\AA, is somewhat
similar to the {\it GALEX}  $NUV$, which has response to the $\lambda$ range
1750\AA\ to 2750\AA\ with an effective wavelength of 2267\AA.
The observing times and exposure times are
listed in Table\,\ref{obs_log}.
The OM data were processed with the XMMSAS task {\it omichain}. 
During the course of the observations a 3\arcsec\ southward drift in 
declination
occurred and the OM onboard software did not correct for it. Instead, by using
foreground stars in the Guide Star Catalog or in the 2MASS image or in 
the {\it Spitzer} IRAC 1 (3.6 $\mu$m) image as standard 
stars,
we applied a plate solution to the ultraviolet images to register them to
the same coordinates as the Spitzer and radio continuum images. The task
{\it omichain} creates for each exposure a source list containing 
raw and corrected counts s$^{-1}$
and magnitudes. We used the source-list data on foreground stars in
the field outside of the galaxies to convert image units to counts s$^{-1}$ 
and magnitudes in the
final stacked image and to check on corrections for dead time and
sensitivity degradation.

\subsection{Radio observations at $\lambda 6$ cm}

With the VLA we observed NGC 2207/IC 2163 in the radio continuum 
at a central frequency of
4860.1 MHz for 92 minutes (on the target) in 
B configuration on 2001 April 14 and
for 50 minutes (on the target) in D configuration
on 1995 May 13. The observations were made with one IF pair at 4885.1 MHz 
with a 50 MHz bandwidth and 
the other at 4835.1 MHz with a 50 MHz bandwidth.
The phase center was RA, Dec(2000) = 06 16 22.665, -21 22 06.87. 
For the B configuration (high resolution)
observations, the phase calibrator was 0606-223,
the flux calibrators were 3C 286 and 3C 147, and the polarization calibrators
were 3C 138 and 3C 286. No significant polarization was detected.
 For the D configuration (low resolution)
observations, the phase calibrator was 0607-157 and the
flux standard was 3C 286. Our D configuration observations were not 
appropriate for a polarization calibration.

The AIPS software package was used for the data reduction. After calibrating
the uv data from each of the VLA configurations separately and checking the
separate maps, we combined the uv data sets from the two configurations and
ran the AIPS task IMAGR with ROBUST =$-2$ to make and clean a map with a
synthesized beam of 2.48\arcsec $\times$ 1.3\arcsec\  (HPBW), 
BPA = 8\arcdeg. After convolution
to a circular beam of 2.5\arcsec\  (HPBW) and correction for primary beam 
attenuation, this became our final $\lambda 6$ cm high resolution image.
In the field of interest, the maximum correction for primary beam attenuation
was a factor of 1.2.
In this image, which is displayed in Figure\,\ref{n2207_6cm}, 
a surface brightness of  1 \mJybeam\ corresponds to
$T_b$ = 8.279 K and the rms noise is 0.016 \mJybeam, equivalent to 
$T_b$ = 0.13 K. We find a total flux density from the galaxy pair in this image
S$_\nu$(4.86 GHz) = $0.132 \pm 0.001$ Jy, with about 20\% of this from IC 2163.
The single-dish observations of  
the Parkes-MIT-NRAO survey\citep{griffith94}
list S$_\nu$(4.85 GHz) = $0.10 \pm 0.01$ Jy for NGC 2207; it is not clear 
whether the latter includes IC 2163.

\subsection{Additional Data}

Other images of this galaxy pair that we use here are the WFPC2 {\it HST} 
$B$-band image from \citet{elmegreen00}, the {\it Spitzer} IRAC and MIPS images
from \citet{elmegreen06}, the H$\alpha$ image from \citet{elmegreen01}, the
VLA \HI\ and line-free $\lambda 20$ cm radio continuum images from
\citet{elmegreen95a}, and a radio continuum image at 8.46 GHz 
($\lambda 3.5$ cm) from
the VLA public archives\footnote{Image credit: NRAO/VLA Archive Survey, 
(c) 2005--2007 AUI/NRAO} (Program AK 509)  from 2003 January 14 observations.
Table\,\ref{psf} lists the FWHM of the point-spread functions (PSF) of the
images we use.
The VLA \HI\ and radio continuum images have
Gaussian synthesized beams (point-spread functions). The other images do not,
and some of the non-radio images, such as the {\it Spitzer} 8 \micron\
and 24 \micron\ images and the \xmm\ X-ray images, have significant 
sidelobes.   

\section{\label{overview} Overview of The System}

\begin{figure}
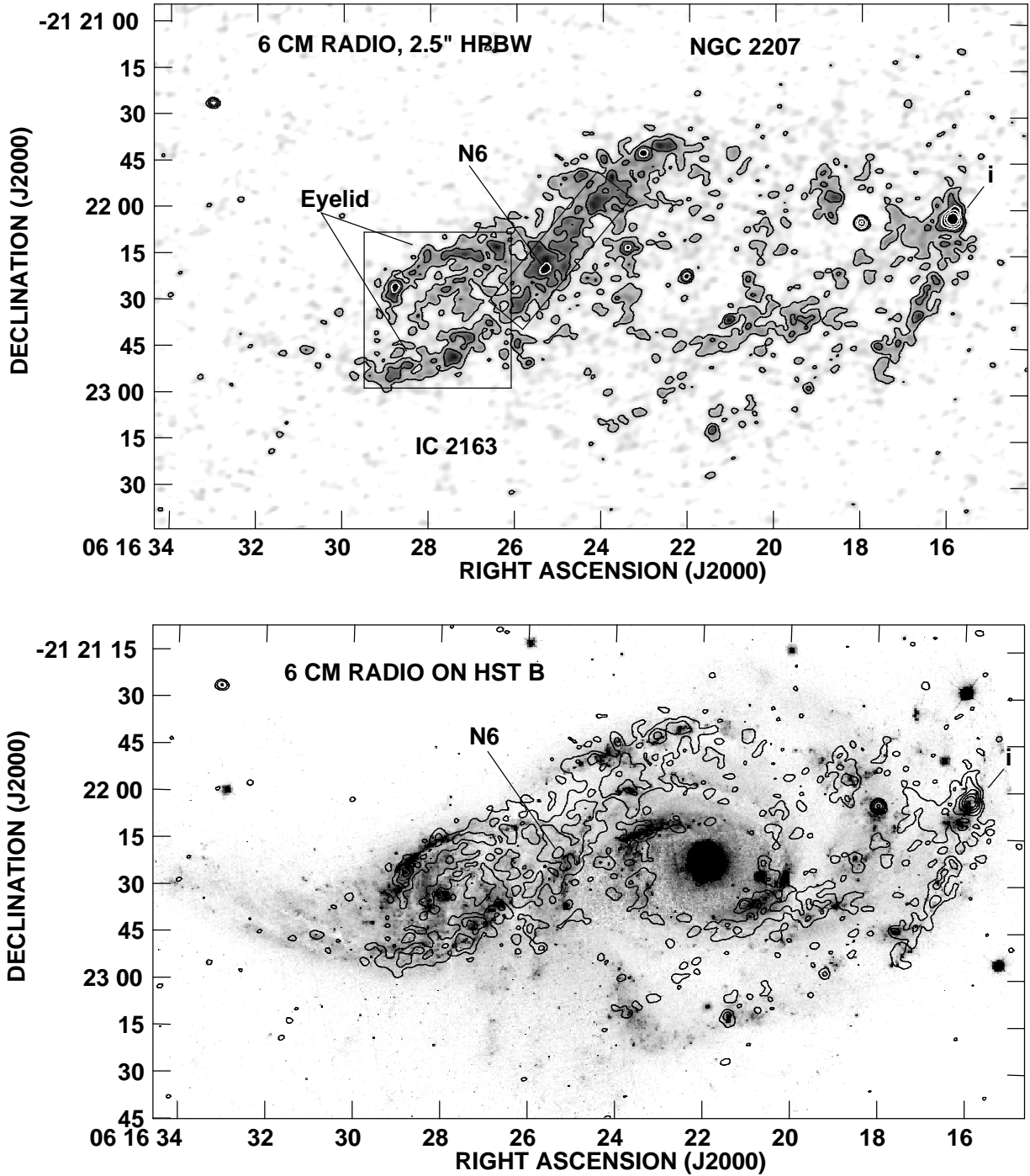

\epsscale{1.1}
\plotone{xmm2207_f1top.ps}
\plotone{xmm2207_f1bot.ps}
\caption{\label{n2207_6cm} Top: Grayscale plus contour display of the \sixcm\
radio continuum image. The rms noise is 0.016
\mJybeam, equivalent to $T_b$ = 0.13 K, and the contours are at 4, 8, 16, 32,
and 64 times the rms noise. The label {\it i} marks {\it feature i}, the 
long tilted box marks the location of the {\it NE radio ridge}, and the label
{\it N6}, the location of massive H I cloud N6.
Bottom: the same contours of \sixcm\ emission 
overlaid on the {\it HST} $B$-band image. 
}
\end{figure}

\begin{figure}
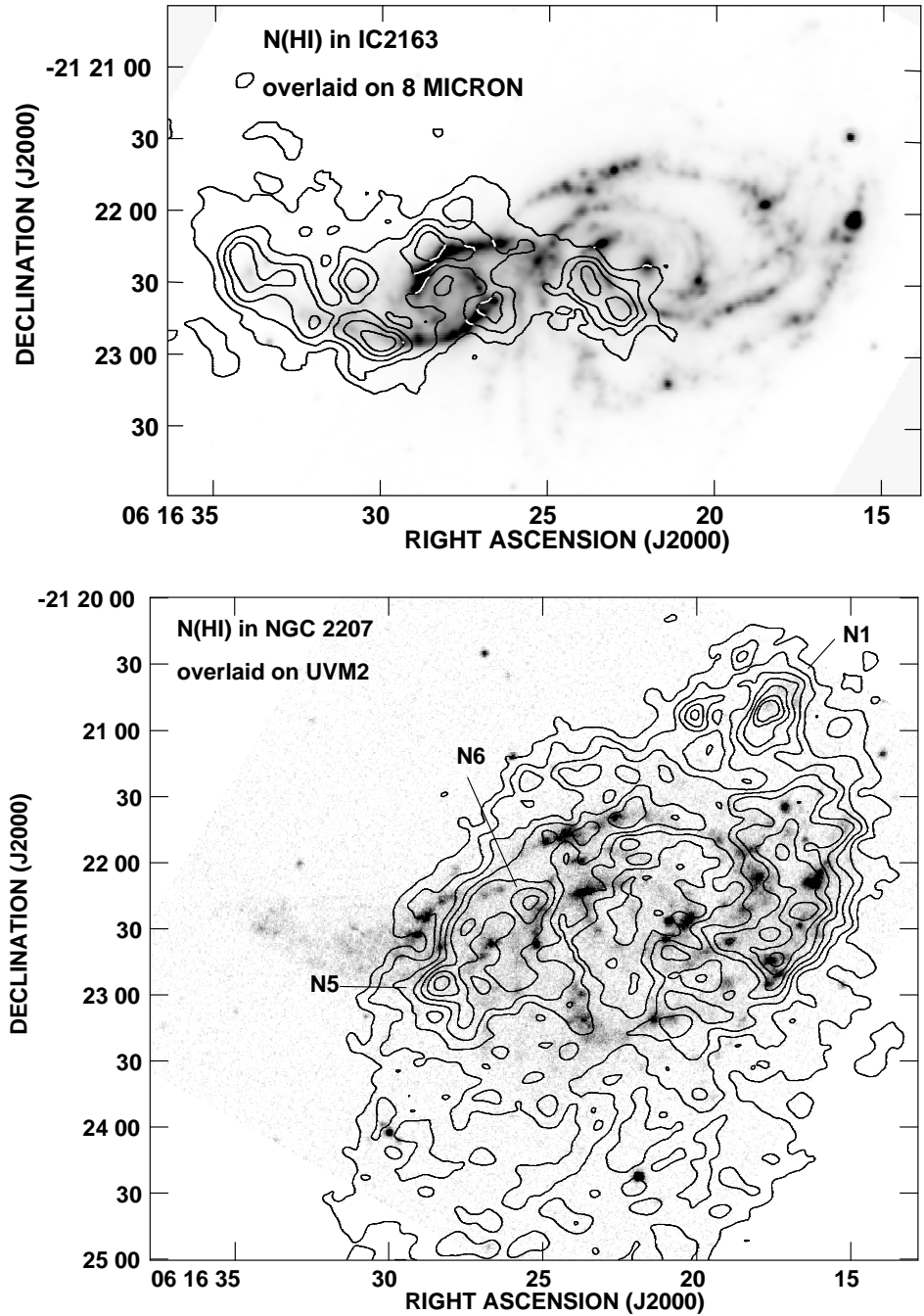

\epsscale{0.85}
\plotone{xmm2207_f2top.ps}
\plotone{xmm2207_f2bot.ps}
\caption{\label{NHI} Top: $N$(HI) contours of IC 2163  
overlaid on the {\it Spitzer} 
8 \micron\ image of NGC 2207/IC 2163. 
Bottom: $N$(HI)
contours of NGC 2207 overlaid on the $UVM2$ image of the galaxy pair.
Contour levels are
$1.25 \times 10^{20}$ \atc\ times 5, 10, 15, 20, 25, 30, and 35. 
 Massive H I clouds N1, N5, and N6 are three of the six 
massive H I clouds in NGC 2207.
}
\end{figure}

Figure\,\ref{n2207_6cm} displays our \sixcm\ radio continuum image and the
WFPC2 {\it HST} $B$-band image  of this galaxy pair, and 
Figure\,\ref{NHI} displays our $UVM2$ image and the Spitzer 8 \micron\ image
 overlaid with contours of the 
line-of-sight \HI\ column density $N(\rm HI)$ associated with each galaxy. 
The two galaxies partially overlap
in projection, with NGC 2207 in front. IC 2163  has an eye-shaped oval midway
out in the disk and a tidal tail on the anti-companion side. The \HI\ image of
IC 2163 reveals a symmetric tidal bridge arm on the companion side; it is
harder to discern in the infrared or optical because it lies behind the central
disk of NGC 2207 but can be faintly traced in the {\it HST} $B$-band
\citep[see Fig. 3 in][]{elmegreen01} and {\it Spitzer} IRAC images.
 The rim of the eye-shaped oval (which we call the 
{\it eyelids}) in IC 2163 is outlined by optical, radio continuum, \HI\
and infrared emission and is 
particularly bright in the {\it Spitzer} 8 \micron\ and 24 \micron\ images.
 The eyelids are one of the large scale shock
fronts that we investigate in Section 5. They are produced by radial streaming
and convergence of orbits due to tidal
forces 
\citep[see the velocity vectors of the model displayed in Fig. 2 of]
[]{elmegreen00}, which concentrate old stars as well as
young stars in the eyelids \citep{elmegreen95a}. The other large scale shock
is the long ridge of enhanced radio continuum emission on the 
companion (eastern) 
side of NGC 2207; in Figure\,\ref{n2207_6cm} a tilted 
box $54\arcsec$ (9 kpc) long is drawn around it. We call it 
the {\it NE radio ridge}. Aside from the highly luminous
{\it feature i}, 
the companion side of NGC 2207 is substantially brighter in the radio continuum
than its anticompanion side, 
with the  brightest large-scale radio emission coming from the  
{\it NE radio ridge}. The radio continuum emission from the adjacent
companion side of IC 2163 is also enhanced. This is evidence of either
disk or halo scraping between the two galaxies.

The {\it NE radio ridge} contains  
optical spiral arms of NGC 2207 visible in the {\it HST} $B$-band image. These 
 are not the outermost spiral arms of NGC 2207 on the
companion side. The bottom panel in Figure\,\ref{NHI} displays an 
outer \HI\ arm of NGC 2207
cutting in front of the western part of the northern eyelid of IC 2163 and 
the eastern part of the southern eyelid. The extinction
due to this outer arm of NGC 2207, which is seen backlit by
IC 2163 in the {\it HST} image \citep{elmegreen00}, partly explains
the faintness of the western part of the northern eyelid and
the eastern part of the southern eyelid in the $UVM2$ image. 
This arm has values of 
N(HI) $\geq 3 \times 10^{21}$ \atc. Assuming the outer disk of NGC 2207
is somewhat metal poor, \citet{elmegreen01} adopt for the relation
between extinction and \HI\ column density,  
$A_v = (0.35 \pm 0.18) \times 10^{-21} \, N$(HI).  
For a foreground dust screen, the ultraviolet extinction $A(UVM2)$ = 
2.6 $ A_v$, and $A(UVW1)$ = 2.0 $A_v$ 
\citep{savage79}. Thus $A(UVM2)$ on this arm
$\geq 2.7 \pm 1.4$ mag.  
In the \sixcm\ image in Figure\,\ref{n2207_6cm}, another outer arm 
of NGC 2207 is visible cutting across just north of the nucleus of IC 2163. 

There is also significant gas 
in the eyelids of IC 2163.
The top panel of Figure\,\ref{NHI} displays the N(HI) associated with IC 2163.
Using the Swedish ESO Submillimeter Telescope (SEST),
 \citet{thomasson04} detected 
\CO\ emission from both
disks; the brightest \CO\ emission is from the central part of
IC 2163 and has an integrated intensity, averaged over the beam, of 6 K \kms,
equivalent to $10^{21}$ H$_2$ cm$^{-2}$ if we use the Milky Way
conversion factor $X_{\rm CO} = N({\rm H}_2)/I_{\rm CO}$ = 
$1.8 \pm 0.3 \times 10^{20}$ from \citet{dame01}. 
The resolution of SEST (43\arcsec\ HPBW) is too low to tell
if this \CO\ emission is mainly from the {\it eyelids} although the
double-peaked nature of the \CO\ line profile of IC 2163 suggests
this may be the case \citep{struck05}.
 Except for a bright \CO\ source on the northwest inner arm of 
NGC 2207, the SEST \CO\ image closely resembles
the {\it Spitzer} MIPS image at 160 \micron\ of this galaxy pair 
\citep{elmegreen06}.
Both have about the same resolution, and the close correspondence tells us
that the cooler dust measured by
the 160 \micron\ emission has about the same distribution globally 
as the molecular gas, which is not surprising. We can infer 
the distribution of
gas from the distributions of cooler and warmer dust, as indexed by the
70 \micron\ and 24 \micron\ emission, respectively. Aside from {\it feature i},
the brightest 70 \micron\ and 24 \micron\ emission in this galaxy pair
 is from the eyelids 
\citep[see the figure in Section 5.1 below and the HiRes deconvolution
of the 70 \micron\ and 24 \micron\ images in][]{velusamy08},
and thus the
highest concentration of gas in IC 2163 is in the eyelids.
 
\begin{figure}
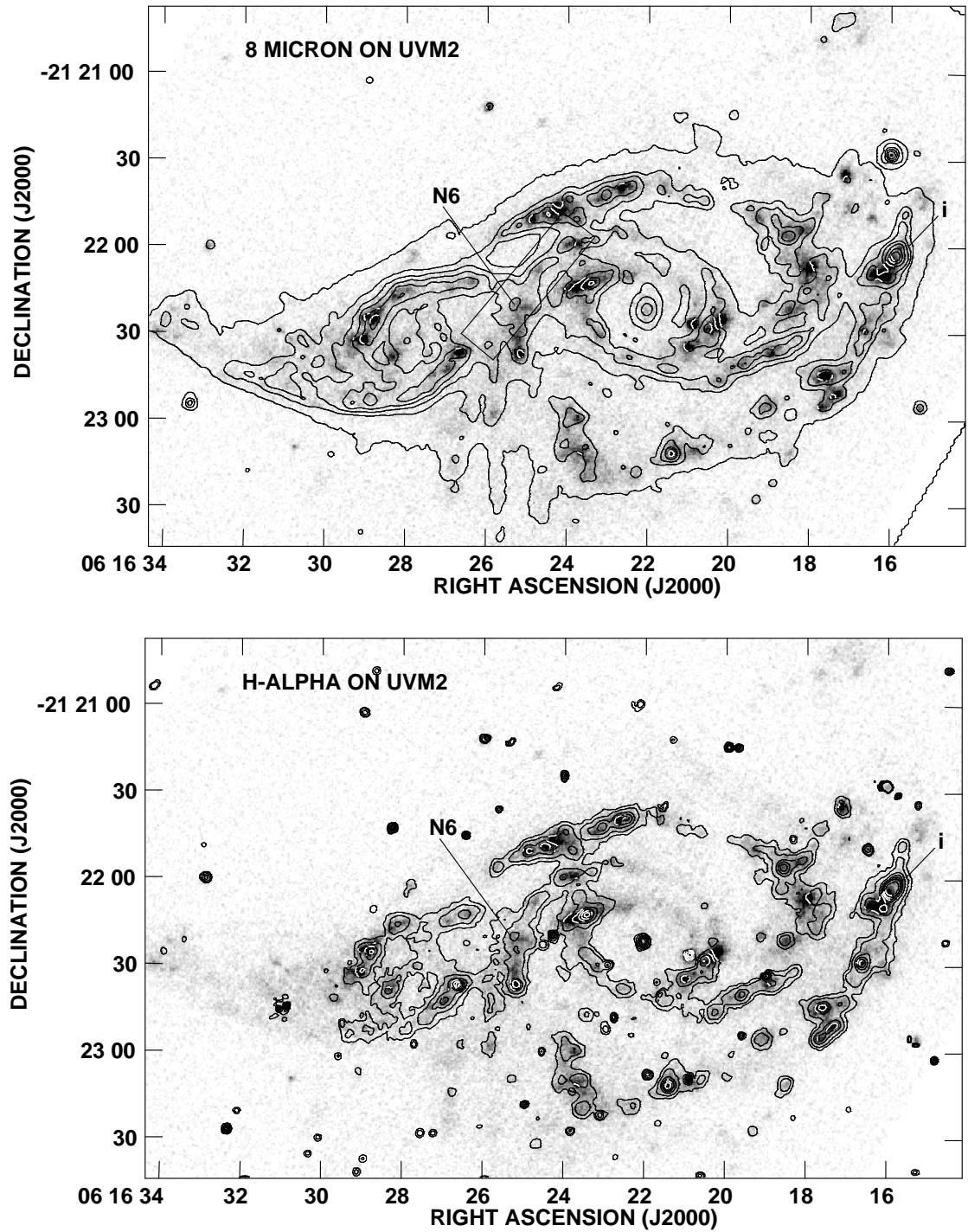

\epsscale{1.0}
\plotone{xmm2207_f3top.ps}
\plotone{xmm2207_f3bot.ps}
\caption{\label{n2207_8m_uvm2} Top: {\it Spitzer} 8 $\mu$m contours overlaid 
on the \xmm-OM $UVM2$ image. The contour levels are 
0.5, 2, 4, 8, 16, 32, and
64 MJy sr$^{-1}$. The long tilted box marks the location of the {\it NE radio
ridge}.  Bottom: Continuum-subtracted H$\alpha$ contours over the $UVM2$ 
image.
}
\end{figure}

\citet{elmegreen95a} identified
11 unusually massive ($10^8 - 10^9$ \msun) \HI\ clouds in NGC 2207/IC 2163.
Most of these are not sites of active star formation.
Clouds N1, N5, and N6 (three of the six massive \HI\ clouds associated 
with NGC 2207) are labelled in  
Figure\,\ref{NHI}. The center of \HI\ Cloud N6 is at 
RA, Dec (2000) = 06 16 25.560, $-21$ 22 19.08. 
The brightest \sixcm\ radio continuum source on
the  {\it NE radio ridge} and the prominent {\it Spitzer} infrared clump 
IR 12 (see Section 4) are $3''$ west, $1''$
south of the center of \HI\ cloud N6; 
SN 2003H is $2''$ east, $5''$ south  of the
center of \HI\ Cloud N6. 
\HI\ cloud N5 obscures the ultraviolet emission from the eastern part of
the southern eyelid of IC 2163. At the center of Cloud N5, N(HI) =
$4.8 \times 10^{21}$ \atc, which corresponds to an A($UVM2$) of 
$4.4 \pm 2.2$ mag.   
\citet{elmegreen93} discuss
the formation of massive \HI\ clouds and tidal dwarf galaxies by 
large-scale gravitational 
instabilities in the gas and suggest that \HI\ Cloud N1 in the outer part of 
NGC 2207 where large $z$ motions are creating a warp may be in the 
process of
forming a tidal dwarf galaxy. The only stellar emission detected from Cloud
N1 forms a bow-shaped arc in the northwestern part of the cloud visible,
for example, in the Digitized Sky Survey image, 
in the blue-band plate in \citet{elmegreen95a},
and in the $UVM2$ image in Figure\,\ref{NHI} and
Figure\,\ref{n2207_8m_uvm2}. It appears that star formation has commenced in
only this part of Cloud N1.

\section{\label{irclumps} Star-Forming Clumps}

Figure\,\ref{n2207_8m_uvm2} displays the $UVM2$ image overlaid with 
{\it Spitzer} 8 \micron\ contours and with H$\alpha$ contours,
respectively. Aside from the
effects of extinction, this demonstrates the generally good correspondence
between these three tracers of recent star formation. 
We use photometry in \sixcm\ radio continuum, 8 \micron, $UVM2$, 
$UVW1$, and H$\alpha$ bands to study star-forming clumps in this galaxy pair
at a resolution of 2.0\arcsec\ -- 2.5\arcsec (0.3 to 0.4 kpc).

\begin{figure}
\epsscale{0.9}
\plotone{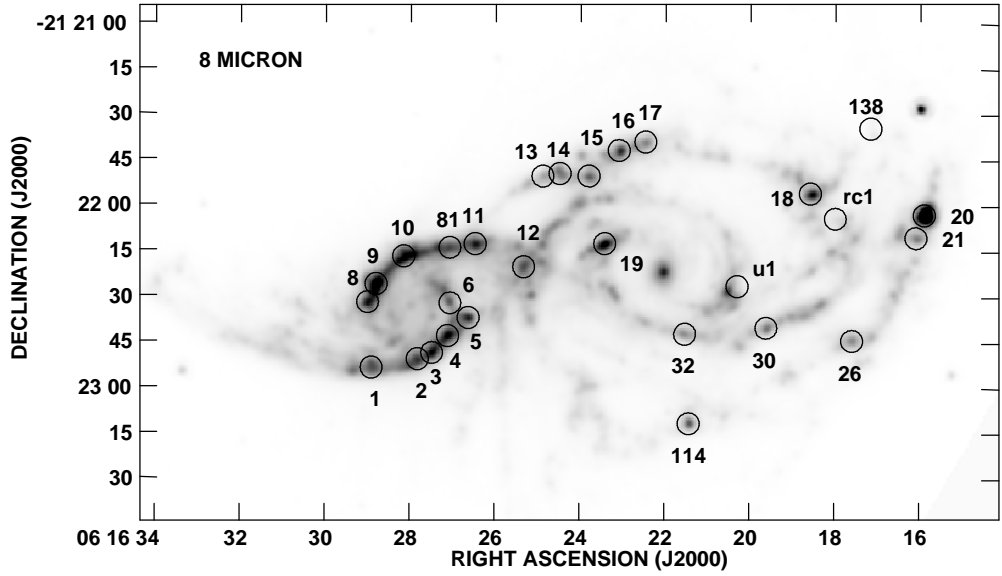}
\caption{\label{clumps} {\it Spitzer} 8 \micron\ image with numbers 
from Table 2 labelling the
measured clumps. The 3.6\arcsec\ radius of each circle 
is the aperture
radius.   
}
\end{figure}

\citet{elmegreen06} did photometry in the Spitzer IRAC bands of
225 bright clumps in this galaxy pair by  
using {\it phot} in IRAF with an aperture radius of 3.6\arcsec\ = 0.61 kpc,
 and a local background
annulus with inner radius = 9.6\arcsec\ and outer radius =
15.6\arcsec\ concentric with the source aperture. The clumps are star 
complexes. In general, the source aperture
includes collections of OB associations and older star clusters 
\citep[see examples in][]{elmegreen06}. 
After registering the 8 \micron\ and ultraviolet images to
the same coordinate grid as the high resolution $\lambda 6$ cm image,
we chose 28 prominent clumps in the 8 \micron\  and/or
$UVM2$ images; these are labelled in Figure\,\ref{clumps}, which displays the
8 \micron\ emission. 
For each clump, Table\,\ref{n2207clumps} lists the 8 \micron\
flux density from \citet{elmegreen06}, the H$\alpha$ flux from 
\citet{elmegreen01}, and the $\lambda 6$ cm flux density, $UVM2$ magnitude,
and $UVM2 - UVW1$ color measured with the same choice of source aperture and 
local background annulus as for the IRAC measurements. 
The uncertainty in the \sixcm\ flux density of each clump is
0.04 to 0.05 mJy. In addition to
free-free radio emission from the \HII\ regions, the source aperture is likely
to include nonthermal radio emission from the spiral arms, some of which is
removed by the local background subtraction. 
For $UVM2$ and $UVW1$ we took from the \xmm\ User's Handbook the zero-points
for the magnitudes (defined such that Vega = 0.025 mag) and the conversion
factors to get from the count rates to flux densities in mJy.
The conversion factors 
are for white dwarfs and thus the values of the $UVM2$ flux density used in
Table\,\ref{n2207clumps} for the ratio of 
8 \micron\ to $UVM2$ flux density are rough estimates.    

The numbering of the clumps is the same as 
in \citet{elmegreen06} except for the added clumps $u1$  
at RA, Dec (2000) = 06 16 20.298, -21 22 26.47 and
$rc1$ at RA, Dec (2000) = 06 16 17.996, -21 22 04.16. 
In section 6, we find that clumps IR 11, IR 21, and rc1 coincide
with discrete X-ray sources. The first eleven clumps in 
Table\,\ref{n2207clumps} are in IC 2163; the rest are in NGC 2207.

The values of $A_v$ in Table\,\ref{n2207clumps} for the NGC 2207/IC 2163
clumps are upper limits obtained from the \sixcm\ flux density and 
H$\alpha$ flux for case B recombination 
with $T_e = 10^4$ K by assuming all of the
\sixcm\ flux density after subtracting the local background is 
optically-thin free-free emission.
For a number of the clumps the $A_v$ upper limits are quite large; 
this leads us to suspect that some clumps 
include significant nonthermal radio emission at \sixcm.
Within a given clump, the
extinction is far from uniform as the {\it HST} observations found lots of
dust features on very small spatial scales near star clusters whose blue colors
indicate little extinction \citep{elmegreen01}.

As an example of what is expected for the flux density ratio
$S_\nu$(8 \micron)/$S_\nu$(6 cm) of an OB association,
we present data in Table\, \ref{m81} on the  
giant \HII\ regions in M81. We chose the 11 giant radio \HII\ regions 
in M81 which have the highest 
signal-to-noise in the \sixcm\ radio continuum observations of 
\citet{kaufman87} and a radio spectral index $\alpha$ 
consistent with optically-thin free-free emission.
 For these M81 \HII\ regions, 
Table\, \ref{m81} uses the
\sixcm\ radio continuum flux densities  from 
\citet{kaufman87} and the {\it Spitzer} 8 \micron, 24 \micron, and
{\it GALEX}  $NUV$ flux densities from \citet{perez06} and finds
the mean value of the ratio 
$S_\nu$(8 \micron)/$S_\nu$(6 cm) = 19 with the standard deviation 
$\sigma$ of the sample = 5, and the mean value of the
ratio $S_\nu$(24 \micron)/$S_\nu$(6 cm) = 36 with 
$\sigma = 9$. (The average measurement uncertainties are 
$\pm 3$ and $\pm 5$, respectively). 
Comparison of these ratios with global values for entire galaxies is given
in Section 5.3.   
The M81 \HII\ regions have
relatively low extinction. Two estimates
of the extinction $A_v$ are listed for each  \HII\ region: $A_v$(K) from 
\citet{kaufman87} is derived from $S_\nu$(6 cm)/S(H$\alpha$);
$A_v$(PG) from \citet{perez06} is obtained from the line ratios
H$\alpha$/H$\beta$ and H$\alpha$/Pa$\alpha$. Except for one \HII\
region, these two methods give the same values for $A_v$
within the uncertainties of the radio data.

The 8 \micron\ emission from star-forming regions is generally attributed to
young massive stars exciting 
aromatic hydrocarbon (PAH) emission bands or heating very small grains.
Far ultraviolet emission from $B$ stars as well as $O$ stars can produce
significant PAH emission via fluorescence \citep{peeters04}.
 Emission at 8 \micron\ 
depends on mechanisms involved in the formation and destruction of
PAH molecules as well as the local SFR \citep{calzetti05, 
perez06, dale07}.

 From IRAC color-color plots,
\citet{elmegreen06} find that the 8 \micron\ emission of most 
of the IRAC clumps in NGC 2207/IC 2163 is PAH dominated.
  
The mean value of the ratio $S_\nu$(8 \micron)/$S_\nu$(6 cm) for the
fifteen
prominent star-forming clumps in NGC 2207 is 18 with standard deviation
$\sigma$ of the sample = 8. Within the uncertainties, this mean value is
the same as that for  the giant radio \HII\ regions in M81. 
We excluded the clump rc1 since it is probably either
a radio supernova or a background quasar (see discussion below and in 
Section 6). In contrast, the mean value of this ratio for the ten
prominent clumps on the IC 2163 {\it eyelids} is 34 with $\sigma = 16$.
The clumps IR 5, IR 8, IR 10 on the eyelids and
 IR 6 on the inner spiral arm of IC 2163 
have values of $S_\nu$(8 \micron)/$S_\nu$(6 cm) $\geq 40$. 
The clumps on the {\it eyelids} are more luminous at 8 \micron\ but have
the same mean $S_\nu$(6 cm) as
the clumps in NGC 2207 (aside from {\it feature i}). As the ten 
{\it eyelid} clumps are detected in H$\alpha$, they contain $OB$
associations, but they may in addition contain somewhat older star clusters
with ages up to 100 million years and thus
could be dominated at 8 \micron\ by PAH excitation involving $B$ stars rather
than $O$ stars. Section 5 mentions another possible 
explanation for the high values of this ratio on the {\it eyelids}. 
 
Although the source aperture used in Table\, \ref{n2207clumps}  to measure the
clumps is much greater in linear diameter (1.2 kpc vs
300 pc) than for the measurements of the M81 giant \HII\ regions and
thus may include emission, such as nonthermal radio emission from the
spiral arm, unrelated to the OB associations, 
a number of clumps in NGC 2207/IC 2163 have a value
of the 8 \micron\ to \sixcm\ flux density ratio similar to those of
the M81 giant \HII\ regions. Either subtraction of the local background has
sufficiently removed the emission unrelated to the OB associations, or
these are examples at 8 \micron\ analogous to
 the usual FIR-radio continuum correlation for galaxies (see Section 5.3)

The three brightest discrete \sixcm\ sources in NGC 2207/IC 2163 are
the clumps IR 20 (which is {\it feature i}), rc1, and IR 12 in 
massive \HI\ cloud N6. All three have 
low values of the 8 \micron\ to \sixcm\ flux density ratio compared to the
M81 giant \HII\ regions.
Nonthermal radio emission is significant in these three
sources. {\it Feature i} is the most luminous
radio continuum source and 
contains the most luminous H$\alpha$ region in the galaxy pair.
VLA snapshots at \sixcm\ and $\lambda 20$ cm by \citet{vila90} 
indicate that  {\it feature i} is dominated by nonthermal radio emission, 
with a radio spectral index $\alpha$ (where $S_\nu \propto \nu^{-\alpha}$)
ranging from 
0.7  in its 1\arcsec\ radio core to 0.9 averaged over a $7.5'' \times 7.5''$ 
box. Clump rc1, the second brightest discrete radio
continuum source in Figure\,\ref{n2207_6cm}, is unresolved in that
image. Clump rc1 is reasonably bright in
$UVM2$ but does not appear as a significant clump in H$\alpha$. The
faintness in H$\alpha$ is not the result of extinction, since rc1 is also
rather faint at 8 \micron. Thus rc1 is a nonthermal radio source. It may be
a supernova remnant, a radio supernova,  or a background quasar. Because we 
detect rc1 as an X-ray source, we defer further discussion of it to 
Section 6. Clump IR 12 lies in the region of brightest nonthermal radio
emission on the {\it NE radio ridge}. 

Table\,\ref{n2207clumps}
also lists the flux density ratio $S_\nu$(8 \micron)/$S_\nu (UVM2)$,
which is sensitive to the amount of extinction, to the distribution of 
extinction, to age, and to star-formation history. 
For the star-forming clumps that are bright in the ultraviolet,
low values of this ratio indicate relatively low extinction. 
Very high values of this ratio may indicate
lots of dust and considerable absorption.
From the values of the
$S_\nu$(8 \micron)/$S_\nu (UVM2)$ ratio, it appears
that clumps  IR 21, IR 26, u1, 
IR 138, and rc1 have relatively low extinction (e.g., values of $A_v$
similar to the M81 \HII\ regions in Table\, \ref{m81}),
and that most of the clumps
on the {\it eyelids} and {\it feature i} suffer high extinction.
The four brightest clumps in $UVM2$ are u1, IR 21, IR 19, and IR 26.
Relative to the H$\alpha$ and radio continuum emission, the ultraviolet
emission from u1 is displaced toward the outer edge of the arm.
IR 138 is a bright $UVM2$ clump that is faint in  H$\alpha$ and 8 \micron\
emission and not detected at \sixcm. On
the {\it HST} image, 
it coincides with a short, thin string of sources in the NW
interarm of NGC 2207. Clumps u1 and IR 138 are
probably slightly older star complexes with little dust and with ultraviolet
emission mainly from $B$ stars rather than $O$ stars.
SN 1999ec at RA, DEC (2000)= 06 16 16.18, -21 22 10.1 \citep{vandyk03} is
1.2\arcsec\ E, 0.3\arcsec\ S of the center of IR 21.
IR 21 could be bright in the 
ultraviolet due to shock excitation of circumstellar gas by SN 1999ec
(see comments about the corresponding X-ray source in Section 8).
IR 21 resembles IR 19 and IR 26 in being fairly bright in H$\alpha$, 
particularly prominent in $UVM2$,
and in having values for the 8 \micron\ to \sixcm\ flux density ratio 
in the same range as the M81 giant radio \HII\ regions.
 A simple interpretation is
that these three clumps are OB associations with less extinction than most
of the other clumps and that the H$\alpha$, and most of the \sixcm\ emission
from IR 21 is from the \HII\ region, not from the Type Ib SN 1999ec.

 For eleven 
of the clumps, the value of the 8 \micron\ to $UVM2$ flux density ratio is
greater than or equal to 120. These include {\it feature i}, eight of 
the ten clumps on the eyelids, and the one clump on the inner spiral of 
IC 2163. It is not surprising to find that {\it feature i} and the eyelid 
clumps 
suffer high extinction. {\it HST} observations \citep{elmegreen06} reveal a
large opaque dust cloud occulting part of {\it feature i}. The eyelids 
contain a large concentration of gas; this is in addition to the extinction
from the outer arm of NGC 2207 cutting in front.

\section{\label{shocks} Large-Scale Shock Fronts}

\begin{figure}
\epsscale{1.0}
\plotone{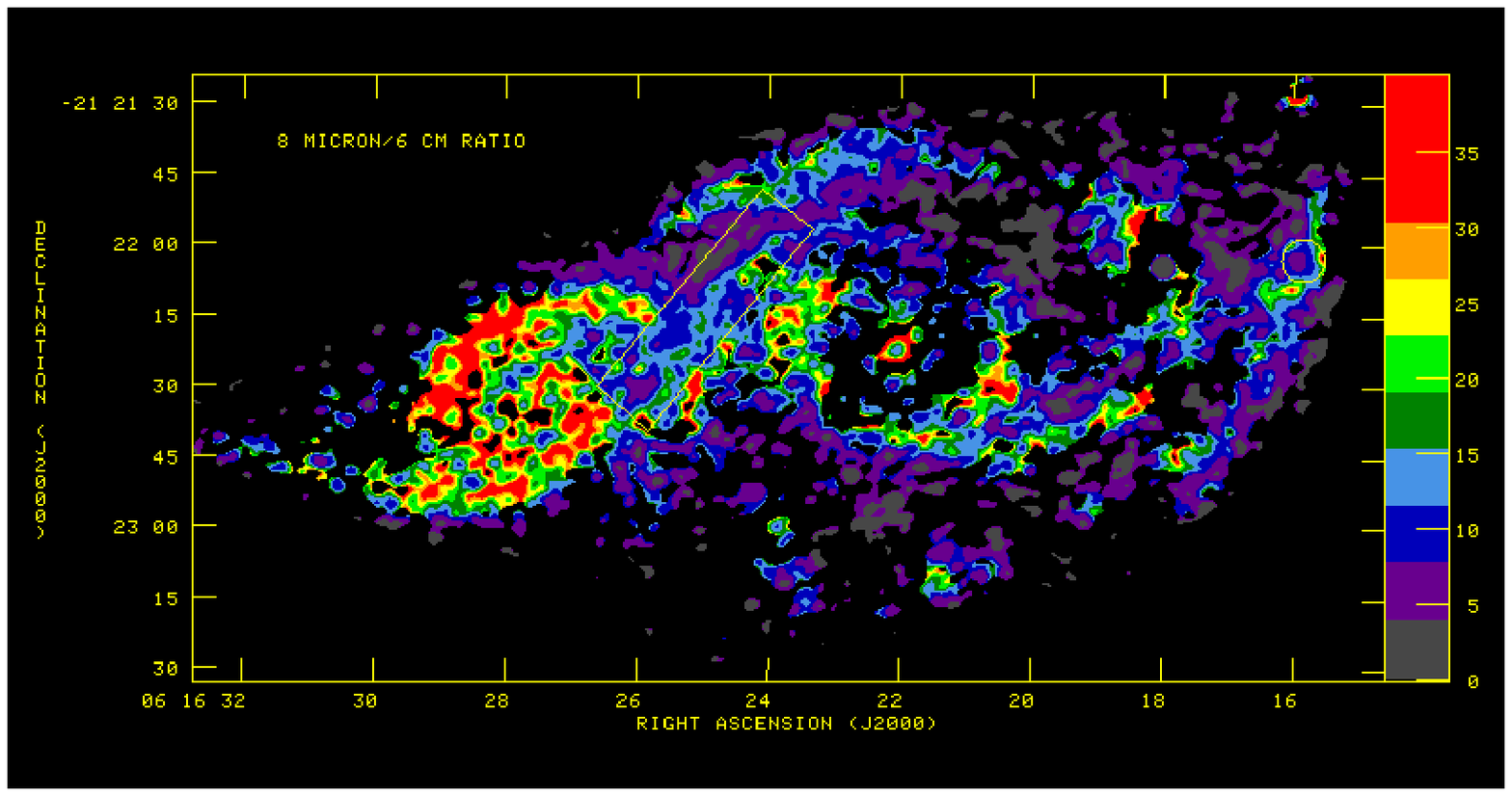}
\plotone{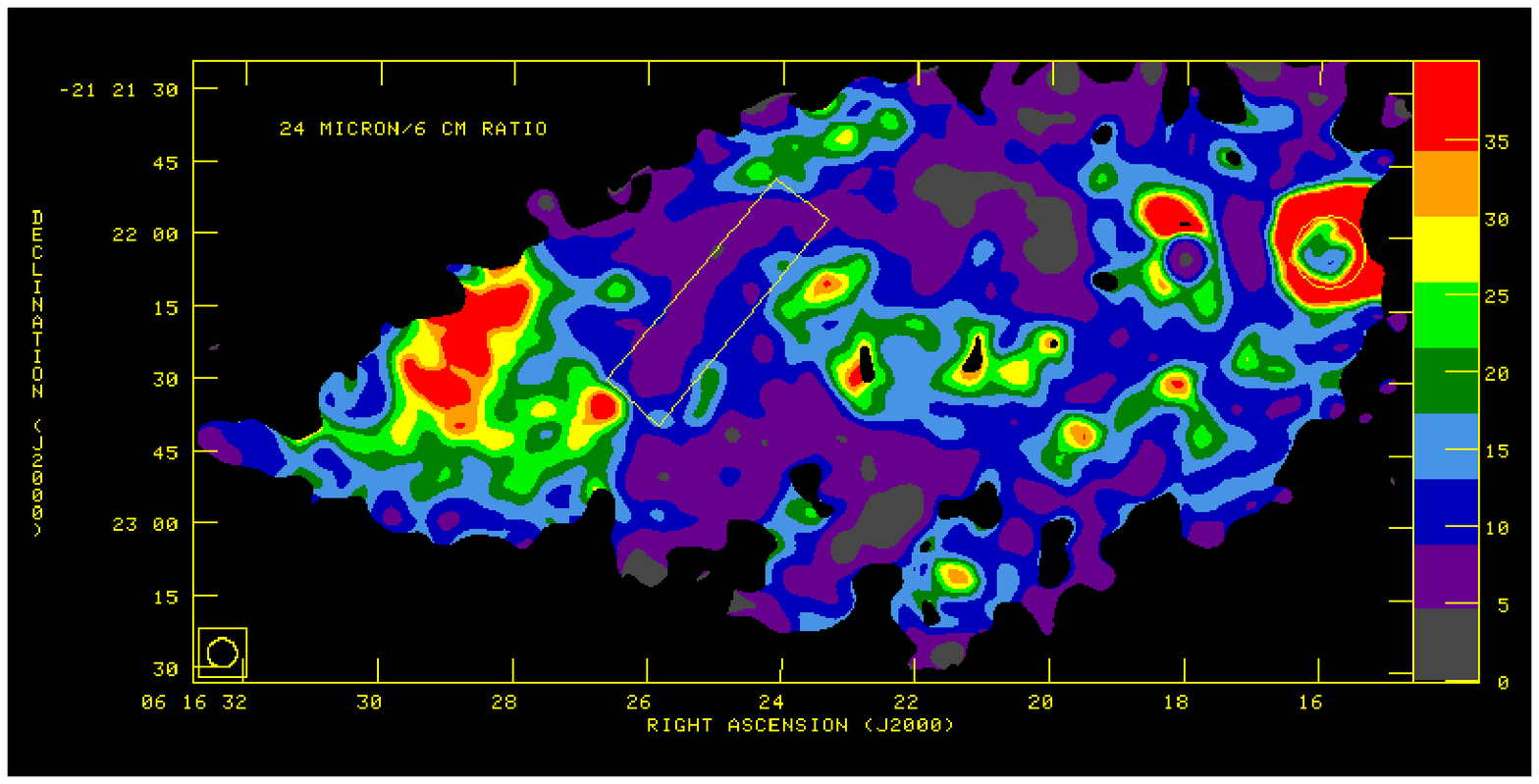}
\caption{\label{n2207_8m_6cm} Top: Surface brightness ratio of 
{\it Spitzer} 8 $\mu$m 
to \sixcm\ radio continuum emission.
Bottom: Surface brightness ratio of {\it Spitzer} 24 \micron\ 
to  \sixcm\ radio continuum emission.
The long tilted box marks  the location of the {\it NE radio ridge}, and 
the circle marks the location of {\it feature i}. For the M81 HII regions,
the mean value of $S_\nu$(8 \micron)/$S_\nu$(6 cm) is $19 \pm 5$,
and the mean value of $S_\nu$(24 \micron)/$S_\nu$(6 cm) is $36 \pm 9$.
In comparison the values of $I_\nu$(8 \micron)/$I_\nu$(6 cm) on the
{\it eyelids} of IC 2163 are high and the values of 
$I_\nu$(24 \micron)/$I_\nu$(6 cm) on the {\it NE radio ridge} are low. 
}
\end{figure}

We consider the two large-scale shock fronts in this system, i.e., the 
{\it eyelids} of the eye-shaped oval in IC 2163 and the {\it NE radio ridge} in
NGC 2207. Figure\,\ref{n2207_8m_6cm} shows the difference between the two
shock fronts. The top panel in this figure displays the 
ratio  of the 8 \micron\ surface brightness $I_\nu$(8 \micron) 
to the \sixcm\ radio continuum surface brightness $I_\nu$(6 cm),
and the bottom panel, the ratio of the  24 \micron\ surface
brightness $I_\nu$(24 \micron)   
to \sixcm\ radio continuum surface brightness. To match the resolution
of the 24 \micron\ image, we made
a \sixcm\ radio continuum image with a synthesized  beam of 6\arcsec\ 
from the \sixcm\ radio observations described in Section 2.2. 

Two pixels in {\it feature i} were hard saturated and thus blanked in the
24 \micron\ Basic Calibrated Data. For these two pixels, we substituted 
twice the 24 \micron\ surface brightness at the FWHM of the point-spread
function.The donut-shaped appearance of {\it feature i} and its environs in the
$I_\nu$(24 \micron)/$I_\nu$(6 cm) ratio image  
(lower panel of Figure\,\ref{n2207_8m_6cm}) results because the radio images
have been cleaned of sidelobes (the diffraction patterns), but the 
{\it Spitzer} images have not.

\begin{figure}
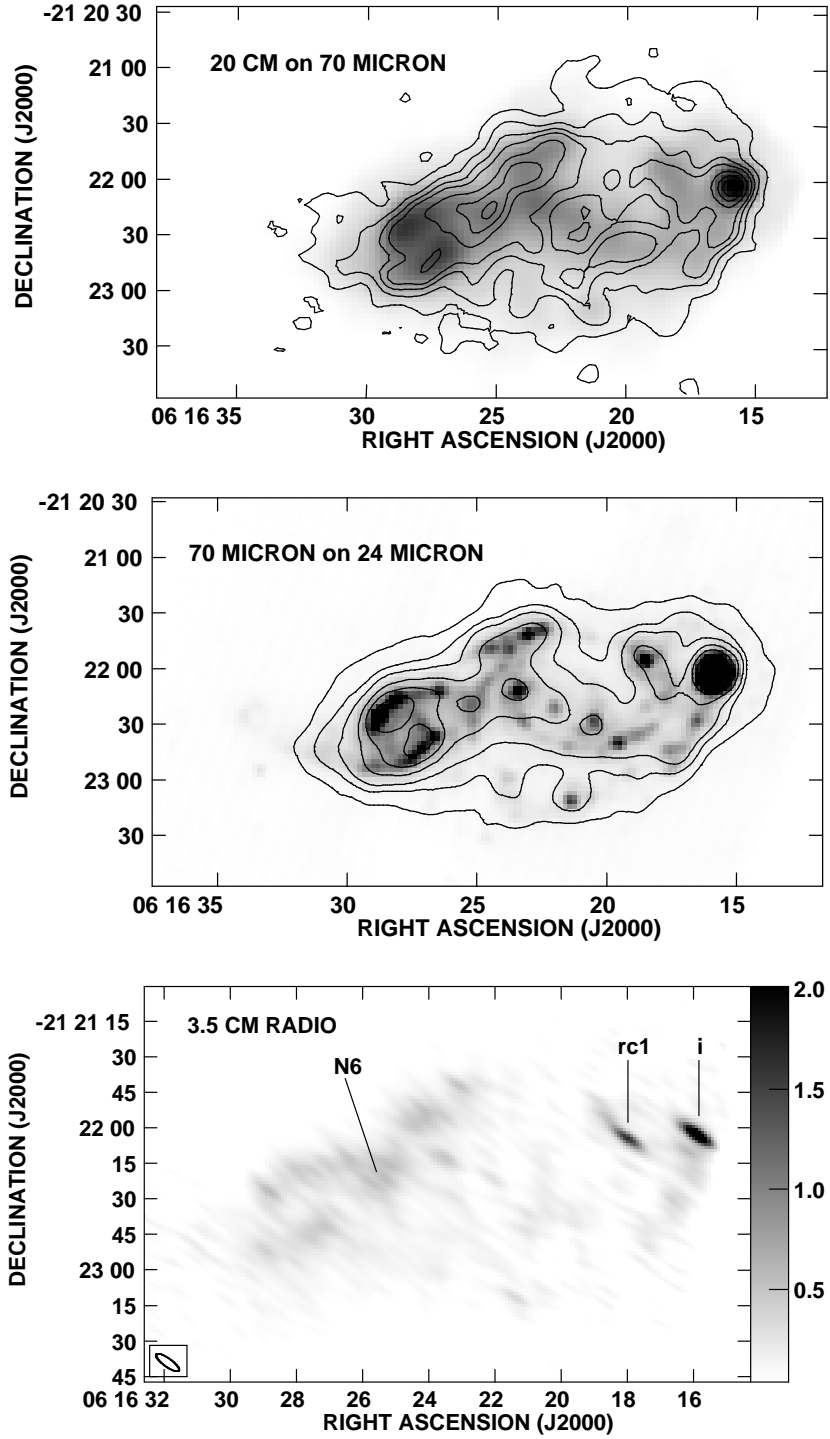

\epsscale{0.75}
\plotone{xmm2207_f6a.ps}
\plotone{xmm2207_f6b.ps}
\plotone{xmm2207_f6c.ps}
\caption{\label{20cm-70micron} Top: Contours of radio continuum emission at
$\lambda 20$ cm (with $13.5'' \times 12''$ resolution)
overlaid on the {\it Spitzer} 70 \micron\ image (with $18''$ resolution).
Contour levels at 1, 2, 3, 4, 6, 8, 10 times 10 K. Middle: Contours of
70 \micron\ emission overlaid on the {\it Spitzer} 24 \micron\ image.
The contour levels are at 20, 30, 40, 60, 80, 100, 120, 140 MJy sr$^{-1}$.
Bottom: Radio continuum emission at $\lambda 3.5$ cm (with 
$11.7''\times 3.5''$ resolution). The eyelids are brighter than 
the {\it NE radio ridge} at 24 \micron\ and 70 \micron, whereas the 
{\it NE radio ridge} is clearly brighter than the eyelids in the
radio continuum at $\lambda 20$ cm. In this
$\lambda 3.5$ cm image from observations in 2003, rc1 is much brighter than 
every source in the galaxy pair except {\it feature i}.  
}
\end{figure}

The differences between the two large-scale shock fronts are also apparent
from the comparison in Figure\,\ref{20cm-70micron} between the {\it Spitzer}
24 \micron\ and 70 \micron\ emission and the radio continuum emission at
$\lambda 20$ cm and $\lambda 3.5$ cm. 
 The {\it eyelids}
are brighter than the {\it NE radio ridge} at 24 \micron\ and
70 \micron\ whereas the {\it NE radio ridge} is appreciably
brighter than the {\it eyelids} at $\lambda 20$ cm.
The value of the ratio of FIR to $\lambda 20$ cm emission varies with
location in the galaxy pair.
The radio continuum emission from NGC 2207/IC 2163 is strongly nonthermal
with a spectral index $\alpha = 0.92$ (where $S_\nu \propto \nu^{-\alpha}$)
\citep{condon83}, and thus nonthermal radio emission makes a 
smaller contribution at $\lambda 3.5$ cm
than at $\lambda 20$ cm. At $\lambda 3.5$ cm, the {\it NE radio ridge} is 
a bit brighter than the {\it eyelids}.

\subsection{The Eyelids}
  
The {\it eyelids} are particularly bright in the 8 \micron\ and 24 \micron\
images \citep{elmegreen06} and also prominent in H$\alpha$ and the radio
continuum (see Figure\,\ref{n2207_8m_uvm2} and Figure\,\ref{n2207_6cm},
resp.). 
 Aside from {\it feature i}, the brightest 70 \micron\ and 24 \micron\
emission in this galaxy pair is from the {\it eyelids} 
(see Figure\,\ref {20cm-70micron}),  and thus cooler dust 
and warmer dust are both strongly concentrated on the {\it eyelids}.
The {\it eyelid} shock front has produced strongly enhanced star formation
in a dusty environment. Emission from warm, very
small grains at 24 \micron\ is considered a better tracer than 
8 \micron\ emission of the SFR \citep{perez06, wu05, calzetti05, dale07}. 
 From Figure\,\ref{n2207_8m_6cm}, one sees that
the values of the ratio 
$I_\nu (8 \, \mu$m)/$I_\nu$(6 cm) on the {\it eyelids} are 
large compared to those of the M81 \HII\ regions and large compared to 
typical values of prominent clumps on the spiral arms of NGC 2207. 
Similarly, along the northern eyelid the values of the ratio
$I_\nu (24 \, \mu$m)/$I_\nu$(6 cm) are greater than 
typical of the spiral arms of NGC 2207, but the effect at both eyelids is
more significant at 8 \micron\ than at 24 \micron.  

The bright 8 \micron\ emission from the eyelids is almost entirely
diffuse emission, not stellar photospheric emission. We checked this by using
Pahre's method \citep{pahre04} to remove the stellar contribution by scaling
the surface brightness of the IRAC 1 (3.6 \micron) image.  

Adolf Witt (private communication) suggests that 
increased collisions between dust grains due to the shock compression
would cause more fragmentation of large grains and increase the number
of small
grains in the size range appropriate for emitting at 8 \micron\ and 
24 \micron. 
Thus collisional shattering of dust grains in the {\it eyelid} shock 
may account for the enhanced 8 \micron\ and 24 \micron\ emission from
the {\it eyelids} by producing a greater  
number of  very small grains with sizes 
appropriate for emitting at 8 \micron\ and 24 \micron\ and by forming more
PAH molecules with sizes $\leq 10^3$ C atoms close to the OB stars. 
According to
\citet{draine07}, it is the PAH particles with sizes $\leq 10^3$ C atoms 
that produce 
most of the emission in the 7.7 \micron\ and 8.6 \micron\ bands.

\subsection{NE Radio Ridge}
 
Aside from {\it feature i}, the brightest large-scale radio continuum 
emission in this galaxy pair comes from the {\it NE radio ridge}.
In H$\alpha$, $UVM2$, 8 \micron, or 24 \micron\ emission 
the {\it NE radio ridge} is no 
brighter than other spiral arms of NGC 2207.
 Figure \,\ref{n2207_8m_6cm} shows that the values of the 
ratio of 8 \micron\ to 
\sixcm\ surface brightness and, particularly, the ratio of 24 \micron\
to \sixcm\ surface brightness are low on the {\it NE radio ridge} 
compared to those of  the M81 \HII\ regions. The
\sixcm\ radio continuum emission is enhanced here without a commensurate 
effect on star formation. The magnetic field $B$ is compressed, 
increasing the synchrotron radio emission, but some condition for enhanced 
star formation is not fulfilled. In IC 2163, as well as NGC 2207, the
companion side is  brighter than the
anticompanion side in the \sixcm\ radio continuum. This is evidence of
disk or halo scraping between the two galaxies. 

The lower panel in Figure\,\ref{n2207_6cm} compares the \sixcm\ radio
continuum image with the {\it HST} $B$-band image. 
South of \HI\ massive cloud N6, the {\it NE radio ridge} includes
two spiral arms of NGC 2207 visible in the $B$-band image, one of
which is backlit by IC 2163, and some emission from IC 2163. North of
cloud N6, the inside edge of the {\it NE radio ridge} coincides with an 
optical spiral arm, but the \sixcm\ radio emission spreads
significantly beyond the outer edge of this arm into the interarm and is
brighter in the interarm.

A question is whether the {\it NE radio ridge} is located in the thin disk of 
NGC 2207 or in a thick disk or in the halo on the back side of NGC 2207 
(relative to us) between the two galaxies.    

If the {\it NE radio ridge} lies in the disk of NGC 2207, it makes sense
to discuss the ridge contrast (defined as the ridge-to-interarm radio disk at
the inside edge of the ridge).  At the 2.5\arcsec\ (425 pc)
resolution of our high resolution \sixcm\ radio continuum image, 
the shock width is
somewhat resolved; going to higher resolution would probably not increase the
ridge contrast by a significant factor. Along the inside edge of the 
{\it NE radio ridge}, the interarm radio disk is detected at a level of
about $2 \times$ the rms noise in this image. 
Along much of this radio ridge from 
PA = 102\arcdeg\ clockwise to 32\arcdeg, the ridge contrast 
in surface brightness is greater than 4 and reaches a maximum of 10 at 
 \HI\ massive cloud N6. In the case of a strong non-cooling
shock transverse to the magnetic field, the shock would increase the
magnetic field $B$ by a factor of 4. 
If equipartition or minimum energy 
or pressure equilibrium of cosmic-ray electrons and magnetic $B$ field applies
\citep[see, for example,][]{beck85}, then the intensity of radio
synchrotron emission
$I_\nu \propto  B^{3 + \alpha(nt)}$,
where $\alpha(nt)$ is the nonthermal radio spectral index. The
galaxy pair has $\alpha$ = 0.92. If
the free-free component of the $\lambda 20$ cm radio continuum emission is
$\leq$ 10\%, then $\alpha(nt)$ = 0.92 to 1.1. With $\alpha(nt)$ in this range
and $I_\nu \propto B^{3 + \alpha(nt)}$, we have the following. (a)  A ridge
contrast in surface brightness of 4 to 10 corresponds to a factor of
1.4 to  1.8 increase in the magnetic field.
(b) A strong magnetohydrodynamic (MHD) shock, which increases $B$ by a factor 
of 4, would increase $I_\nu$ by a factor of 200 to 300. The latter is
ruled out by the observed ridge contrast if the {\it NE radio ridge} is
in the disk. In the limiting case where the shock increases only
the magnetic field and not the cosmic-ray electron density,
$I_\nu \propto B^{1 + \alpha(nt)}$ and a strong MHD shock,
which increases $B$ by a factor of 4, would
increase $I_\nu$ by a factor of 14 to 18 for the above range of
$\alpha(nt)$. This is somewhat greater than the observed
ridge contrast of 4 to 10. This limiting case is more likely to apply in
the halo than in the thin disk.
We conclude that if the
{\it NE radio ridge} is in the disk, it is a broad ridge of somewhat compressed
magnetic field, not a strong MHD shock.

However, if the {\it NE radio ridge} is high off the midplane,
then taking the measured ratio of  the radio surface brightness on the 
{\it NE radio ridge} 
to that of the interarm thin disk at its inside edge does not make sense, and
thus a strong MHD shock may be allowed. The following comparison between
the distributions of \sixcm\ radio continuum emission,
neutral gas, and cool dust in this system provides information relevant
to the question of whether the {\it NE radio ridge} is in the disk. 
In NGC 2207, the distribution 
of 70 \micron\ emission (see Figure\,\ref{20cm-70micron}) and the 
distributions of 160 \micron\ emission \citep{elmegreen06}
and \CO\ emission measured at SEST \citep{thomasson04} generally
correspond with the spiral arms.
This indicates that molecular gas and 
the cool dust are cool fluids in the thin disk.
The \HI\ observations by \citet{elmegreen95a} of this galaxy pair (see also
Figure\,\ref{NHI}) find that along the northern side of NGC 2207, the \HI\
ridge consistently coincides with the spiral arm, but on  the eastern
and western sides of NGC 2207, the \HI\ ridge line often lies in the interarm
region, the massive \HI\ clouds are usually in the  
interarm, and the \HI\ gas has high velocity dispersion.  
The high velocity
dispersion leads \citet{kaufman97} and 
\citet{elmegreen00} to suggest that the \HI\ gas disk may be a few times 
thicker than normal. The \HI\ disk may be flared on the companion and 
anti-companion sides of NGC 2207 to form a thick disk. 
In Figure\,\ref{n2207_6cm} we see that
the bright \sixcm\ radio continuum emission from 
NGC 2207 generally coincides well with the stellar arms 
 except on the {\it  NE radio ridge}. 
The most luminous radio
continuum source on the {\it NE radio ridge} is in the massive \HI\ cloud
N6; cloud N6 and this radio continuum source may be in the thick disk. 
Along the {\it NE radio ridge} from cloud N6 clockwise to position angle
PA = 30\arcdeg, the brightest \sixcm\ radio continuum emission is in the
interarm. The \HI\ emission here is also bright in the interarm. 
The \sixcm\ emission and the \HI\ emission on
the {\it NE radio ridge} appear clumpy.  However, except for cloud N6, the
clumps are not in one-to-one correspondence. More importantly, unlike
the radio continuum, the \HI\ emission from the {\it NE radio ridge} is
no brighter than from the opposite side of NGC 2207. This may be understood
if a substantial fraction  of the radio continuum emission in 
the {\it NE radio ridge} originates in the halo.

On scales greater than about 2 kpc in normal spiral galaxies, 
\citet{adler91} find that the ratio of \CO\ 
intensity to radio continuum surface brightness is
fairly constant. In NGC 2207, the
 70 \micron, 160 \micron, and \CO\
emission tend to be brighter on the eastern side of NGC 2207 than on
its western side (aside from {\it feature i}). 
However, the difference between the
two sides of NGC 2207 is less pronounced for the molecular gas and cool dust
than for the radio continuum. Sensitive \CO\ mapping with higher spatial
resolution than SEST would be useful here.
 
 On the {\it NE radio ridge}, we may be seeing a combination in which the 
bright radio continuum emission at the spiral arm is from the 
thin disk, some of the bright radio emission
is from compressed magnetic fields in a thick disk (e.g., cloud N6),
 but most of the interarm radio emission is from the halo
on the back side of NGC 2207 (relative to us) between the two galaxies.
If the compressed magnetic field is in the halo and there is little 
neutral gas in the halo, then it is easy to understand
why the {\it NE radio ridge} is not a site of extended vigorous star 
formation. 
If the compressed magnetic field is in the thin disk, it seems
necessary to invoke a time delay between compression of the magnetic field
and compression of the neutral gas, which then leads to active star
formation. We note that at the spiral arm, compression due to disk-scraping 
would add to the already existing compression of the spiral density wave 
to produce brighter radio continuum emission. 

The lack of enhanced star formation on the {\it NE radio ridge} is
analogous to the lack of active star formation in most of the massive
\HI\ clouds in this galaxy pair. Much of the \HI\ in the thick disk may
be at too low a volume density and thus there is 
a delay before molecular clouds form.

\subsection{Comments about IR to radio continuum ratios}

From \citet{appleton04} we adopt the notation $q_{{\rm IR}}$ = 
$\log (S_{{\rm IR}}/S_{1.4 \, {\rm GHz}})$, where $S_{{\rm IR}}$ 
is the flux density
in the {\it Spitzer}
8 \micron, 24 \micron, or 70 \micron\ bands or in the {\it IRAS} FIR band, and
$S_{1.4 \, {\rm GHz}}$ is the radio continuum flux density at 1.4 GHz.
In Table\,\ref{q}, we compare the global values of $q_{{\rm IR}}$ for 
NGC 2207/IC 2163 with the median or mean values
for galaxies in the Spitzer First-Look Survey from \citet{appleton04},
for a
sample of 30 or 35 star-forming galaxies in the Spitzer First-Look Survey
from \citet{wu05}, for {\it IRAS} 
galaxies that do not contain a radio-loud AGN from \citet{condon92}, and
for galaxies in the SINGS sample, where we used the data from \citet{dale07}.
For the SINGS sample, we omitted the galaxies with poor quality data that were
excluded by \citet{draine07}, and to have a more suitable comparison with
NGC 2207/IC 2163, we also omitted the nine low metallicity galaxies listed
by \citet{draine07}. Including the  E and S0 galaxies
in the SINGS sample has little effect on the mean values of $q_{{\rm IR}}$
(see Table\,\ref{q}).
For NGC 2207/IC 2163, we used the 1.4 GHz radio continuum flux density
of $393 \pm 9$ mJy from the NRAO/VLA Sky Survey (NVSS) and the {\it Spitzer}
24 \micron\ and 70 \micron\ flux densities from \citet{elmegreen06} but 
revised the 8 \micron\
value to include the aperture correction for extended 
emission\footnote{from http://spider.ipac.caltech.edu/staff/jarrett/irac/}
 and an improved global background subtraction.
The values of $q_{{\rm FIR}}$, $q_{70}$, and $q_8$
for NGC 2207/IC 2163 are consistently below 
the medians or means of the above large samples of galaxies by 2.1 to 
3.2 $\sigma$,  
where $\sigma$ is the standard deviation of the sample. The value
of $q_{24}$ for NGC 2207/IC 2163 is 2 $\sigma$ below the mean values of the
samples in \citet{wu05} or SINGS, but only 1 $\sigma$ below the sample in  
\citet{appleton04}.

These values of $q_{{\rm IR}}$ are for entire galaxies. Table\,\ref{q}
also lists the mean values of $q_{24}$ and $q_8$ for the set of M81
\HII\ regions in Table\, \ref{m81} obtained by using 
 $S_\nu$(20 cm) values from \citet{kaufman87}. The mean values of 
$q_{24}$ and $q_{8}$ for the M81 \HII\ regions are consistent with the
SINGS sample but, respectively, 2.2 
$\sigma$ and 1.8 $\sigma$ greater than those of the sample of
star-forming galaxies in \citet{wu05}. The radio continuum from the
M81 \HII\ regions is optically-thin free-free emission, 
whereas the radio continuum of 
galaxies as a whole is dominated by nonthermal emission. If the galaxies
have a radio spectral index $\alpha$ of 0.8, then  to obtain
$\log [S_\nu({\rm IR})/S_\nu(4.86 \, {\rm GHz})]$, add
$\alpha \log (4.86 \, {\rm GHz} /1.4 \, {\rm GHz})$ to the 
$q_{{\rm IR}}$ values in 
Table\, \ref{q}, i.e., add 0.43 to the mean $q_{{\rm IR}}$ values for the 
galaxy 
samples and 0.05 for the M81 \HII\ regions. The resulting mean values for
the \citet{wu05} sample are
$S_\nu$(8 \micron)/$S_\nu$(6 cm) = 22  versus 19 for the M81 \HII\
regions  and $S_\nu$(24 \micron)/$S_\nu$(6 cm) = 32 versus 36 for 
the M81 \HII\ regions. For the SINGS sample the resulting mean values are
greater than these.
Table\,\ref{totalflux} lists  values of the flux densities at 
\sixcm, 8 \micron, and 24 \micron\
from the {\it NE radio ridge}, from the box drawn around the
{\it eyelids} in Figure\,\ref{n2207_6cm}, and from the galaxy pair
as a whole. The ratio of 8 \micron\ to \sixcm\ radio continuum flux density 
equals 9 for NGC 2207/IC 2163 as a whole and for the {\it NE radio ridge} box.
The ratio of 24 \micron\ to \sixcm\ radio continuum flux density
equals 15 for the galaxy pair as a whole and about 9 for 
the {\it NE radio ridge} box (the latter value is uncertain because the box is
only $14''$ wide). These values are low compared
to  those of the M81 \HII\ regions and the mean values for the galaxies in the 
\citet{wu05} and SINGS samples .
It is clear from Figure\,\ref{n2207_8m_6cm}
and Table\,\ref{totalflux}
that this is the result of excess \sixcm\ radio continuum emission from
large portions of NGC 2207, not just the {\it NE radio ridge} box.
The {\it eyelid} box contains the {\it eyelids} plus the 
region interior to the {\it eyelids}. 
The 8 \micron\ to \sixcm\ flux density ratio = 19 for the 
{\it eyelid} box is appreciably less than most of the values along the
{\it eyelids} because 
 (a) the box includes an outer
arm of NGC 2207 which is bright at \sixcm\ but faint at 8 \micron, and
(b) the outer part of the companion side of IC 2163 has enhanced radio
continuum emission (produced by the same scraping
as the {\it NE radio ridge}).

The flux density ratio $S_\nu$(24 \micron)/$S_\nu$(8 \micron) 
 depends on the strength of the PAH emission. It is sensitive to
variations in PAH formation and destruction \citep{dale07}. 
Values of this ratio
for NGC 2207/IC 2163 as a whole and for the {\it eyelid box} are listed in 
Table\, \ref{totalflux} and are consistent with those listed in
Table \,\ref{q} for the SINGS survey.

\section{\label{results} X-Ray Results}

\begin{figure}
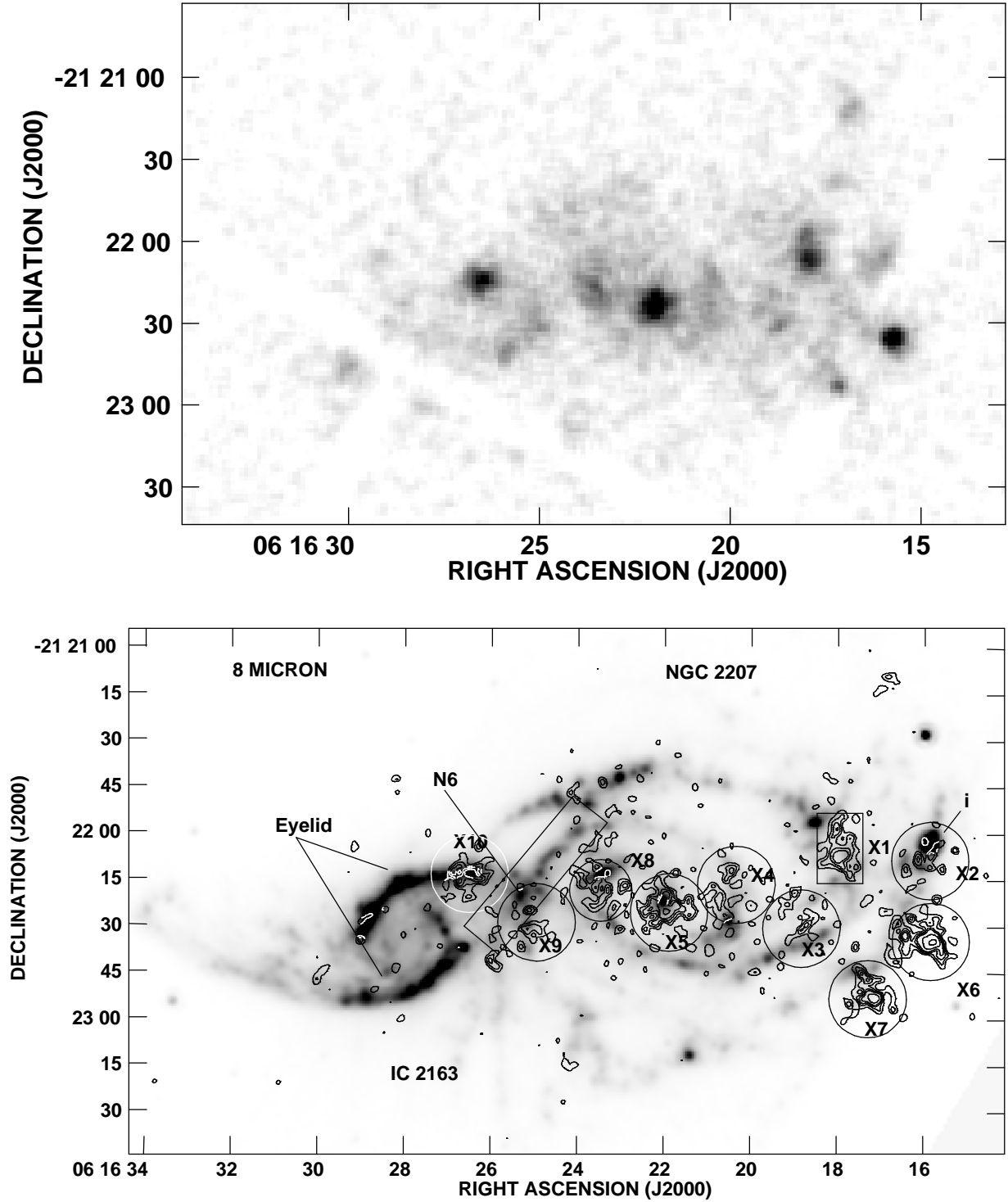

\epsscale{1.1}
\plotone{xmm2207_f7top.ps}
\plotone{xmm2207_f7bot.ps}
\caption{\label{n2207_xray_image} Top: X-ray image of the combined, smoothed, 
screened EPIC pn and MOS data in the 0.5 -- 10.0 keV energy range. The gaps
result from including the pn camera data in this image. 
Bottom: 8 \micron\ {\it Spitzer} image with X-ray contours
from the MOS data only  overlaid. The entire field of interest fits
onto a single CCD chip in the MOS camera data. The sources
refer to Table\,\ref{source_list}. The circles and boxes mark the 
source extraction regions for the X-ray analysis. For the X-ray contours,
the MOS data with FWHM of the PSF $\sim 5''$ was not smoothed, and the
small-scale wiggles in the contours are generally not significant. 
The X-ray image used in the bottom panel has better spatial resolution but 
poorer sensitivity than the X-ray image in the top panel.
}
\end{figure}

The top panel in Figure\,\ref{n2207_xray_image} displays an X-ray image
in the 0.5 -- 10 keV range obtained by combining data from the 
MOS camera with data from the pn camera 
and smoothing the image to get greater sensitivity. The same screening 
criteria as for the X-ray spectra (see Section 2) were applied.     
The bottom panel displays the 8 \micron\ {\it Spitzer} image with 
contours of X-ray emission from the MOS camera data only and with the source 
extraction boxes for the X-ray spectral analysis overlaid. 
The X-ray image used in the bottom panel has better spatial resolution
(FWHM of the PSF $\sim 5\arcsec$) 
than the X-ray image in the top panel but lower sensitivity.
For the discrete X-ray sources,
we subtracted the local background by collecting background
counts from a nearby region with the same area as the source box.
With the MOS cameras, the entire field of interest fits onto a single 
CCD chip. With the pn camera, part of the southern eyelid of IC 2163 and 
parts of the discrete sources X2 and X7 in NGC 2207 fell in a gap between 
two CCDs 

 The X-ray images in Figure\, \ref{n2207_xray_image} 
clearly show the nucleus of 
NGC 2207 and nine other discrete X-ray sources labelled in the figure,
as well as extended X-ray emission.
No X-rays are detected at the position of the nucleus of IC 2163. There
is little activity in the IC 2163 nucleus; it 
is faint (or very faint) in the radio
continuum, ultraviolet, X-rays, 8 \micron, and 24 \micron\ images.
 The nucleus of
NGC 2207 is prominent in  soft X-rays, hard X-rays, radio continuum,
H$\alpha$, $UVW1$, and 8 \micron\ images, but rather faint in the
$UVM2$ image (as a result of extinction). The NGC 2207 nucleus is the only
source in this galaxy pair that is bright in hard X-rays with $E > 5$ keV.

\subsection{The galaxy pair NGC 2207/IC2163}

 The X-ray emission from the discrete X-ray 
sources and the extended emission, which may be
from hot galactic gas, are mainly concentrated in NGC 2207.
Aside from the NGC 2207 nucleus, most of the discrete X-ray sources lie
on the spiral arms. Only a few of these correspond to the prominent IR
or UV clumps discussed in Section 4. 

The encounter models for this galaxy pair \citep{struck05} predicted
soft X-ray emission from diffuse hot plasma at the large-scale shock
fronts. One goal of our \xmm\ observations was to detect such emission.
We find that neither the large-scale shock front along the {\it eyelids} 
nor the {\it NE radio ridge} appears enhanced in extended X-ray emission
relative to the rest of this galaxy pair. X-ray absorption due to the large
concentration of gas in the {\it eyelids} plus gas in the outer arm of 
NGC 2207 cutting in front of IC 2163 may explain why we do not
detect significant, extended, soft X-ray emission from the {\it eyelids}.
Part of the southern eyelid lies in the gap between two CCDs of the pn
camera, but this is also where we find very high extinction in the ultraviolet
for the star forming clumps (Section 4). Most of 
the {\it NE radio ridge} has significantly less extinction than 
the {\it eyelids}. In the models of \citet{struck05}, the
{\it NE radio ridge} is attributed to disk or halo scraping.  
The lack of enhanced X-ray emission from the {\it NE radio ridge} supports
the suggestion that the bright radio continuum emission here arises in a
region of significantly lower density than the thin disk, since no evidence
that a shock due to disk or halo scraping has heated large quantities of gas
to X-ray temperatures. 

\begin{figure}
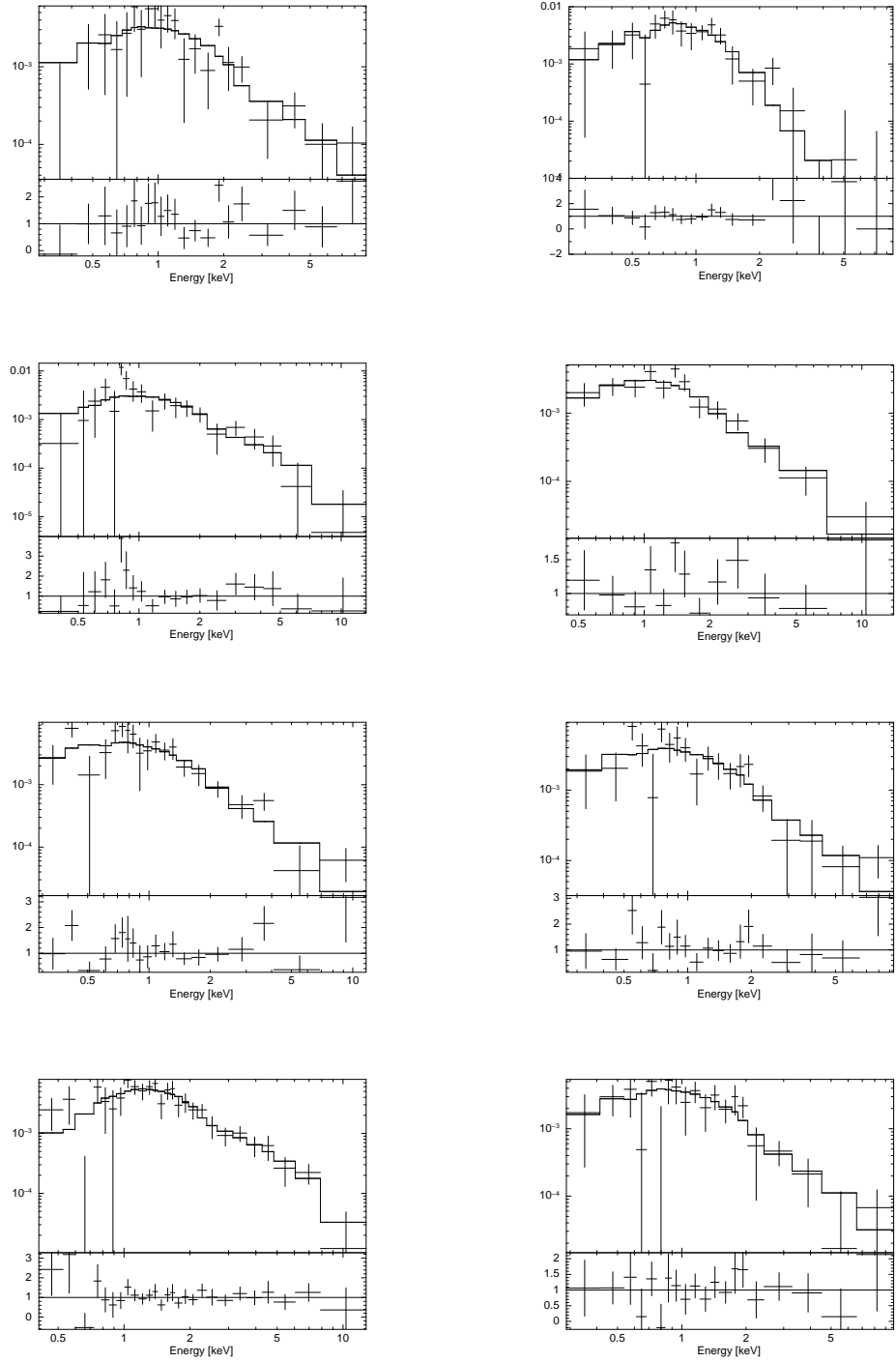

\epsscale{0.5}
\plottwo{xmm2207_f8a.eps}{xmm2207_f8b.eps}

\plottwo{xmm2207_f8c.eps}{xmm2207_f8d.eps}

\plottwo{xmm2207_f8e.eps}{xmm2207_f8f.eps}

\plottwo{xmm2207_f8g.eps}{xmm2207_f8h.eps}

\bigskip
\caption{\label{xray_sources_spec} X-ray EPIC pn spectra of sources labelled
in Figure\,\ref{n2207_xray_image} and
listed in Table\,\ref{source_list}. The spectra were fitted with an absorbed
power law model with Galactic and intrinsic absorption as listed in
Table\,\ref{source_list}. From top to bottom, the spectra in the left column
are for sources X1, X4, X8, and X10 and the spectra in the right column are
for sources X3, X6, X9, and the {\it NE radio ridge}. The ordinate on each 
spectrum is the normalized counts s$^{-1}$ kev$^{-1}$ and the
plot below each spectrum is the ratio of observed-to-model counts.
}
\end{figure}

\begin{figure}
\epsscale{0.5}
\plottwo{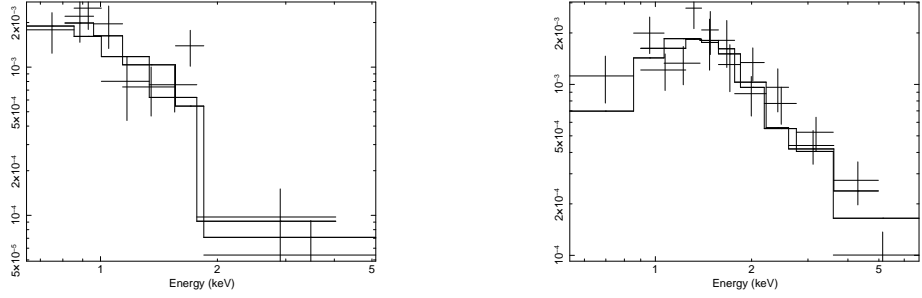}{xmm2207_f9b.eps}
\bigskip
\caption{\label{mos_spec} X-ray MOS spectra of sources X2 (left panel) and 
X7 (right panel). These lie at
the chip edge in the EPIC pn observations.
}
\end{figure}

The X-ray spectra of 
each individual source in Figure\,\ref{n2207_xray_image} plus the 
{\it NE radio ridge} (denoted RR in Table\,\ref{source_list})
were fitted by an absorbed power law model with the absorption column
density by the Milky Way fixed to the Galactic value given by 
\citet{dic90}
and the absorption column density of the  
absorber at the location of NGC 2207 (z=0.00941) left as a free
parameter. The results of these fits are summarized in 
Table\,\ref{source_list}. The EPIC pn X-ray spectra of all of the 
sources except X2, X5, and X7 are displayed in 
Figure\,\ref{xray_sources_spec}. The MOS spectra of sources X2 and 
X7, which lie at  the chip edge in the pn observations,
are displayed in Figure\,\ref{mos_spec}. The pn and MOS spectra of X5
(the nucleus of NGC 2207) are displayed in 
Figure\,\ref{n2207_nucleus_xray_spec}. Most of these X-ray spectra
can be fitted with a power law with an energy spectral index of about \ax=1.0
(equivalent to a photon index $\Gamma$ of 2.0),
except for X2 and X3, which show significantly steeper indices, and X5,
which has a more complicated spectrum. The power-law model for X3
resulted in an unusually steep spectral index 
$\alpha_X$ = $3.7^{+2.6}_{-0.9}$.
A Raymond-Smith plasma is a better representation of the spectrum of X3,
and this is what we list for X3 in Table\,\ref{source_list}.

If the 10 discrete sources in Table\,\ref{source_list} are at the distance
of IC 2163/NGC 2207 and emitting isotropically, then each has an X-ray
luminosity $L_x \geq 2 \times 10^{39}$ erg s$^{-1}$. Thus, each of the nine
discrete sources in the disks of this galaxy pair is a ULX candidate. 
Since each could be a collection of sources, e.g., X-ray binaries, 
rather than a single object, {\it Chandra} high resolution observations
and reobserving to look for variablility are necessary to check on this.
From {\it Chandra} data on 32 nearby galaxies, \citet{colbert04} find
that most of the discrete X-ray sources in the disks of spiral galaxies
have spectra that fit an absorbed power-law with $\Gamma \approx 1 - 2$,
appropriate for high-mass X-ray binaries associated with 
accretion-powered black holes, low-mass X-ray binaries, and 
ULXs and that only the Antennae merger pair
(NGC 4038/39)
contains more than three ULXs. From a study of ULXs in 82 galaxies,
\citet{swartz04} conclude that  14\% of ULX candidates in spiral galaxies are
probably background sources. Only the following four galaxies
in their sample have more than five ULXs: M82, M51, NGC 4038, and NGC 4486.
From \chandra\ observations of the interacting starburst pair NGC 7714/7715,
\cite{smith05} identify 11 candidate ULXs, only two of which are more
luminous than the faintest discrete X-ray source listed in 
Table\, \ref{source_list} for NGC 2207/IC 2163.   
If most of the candidates in NGC 2207 are ULXs, that would be
a greater number than in a typical galaxy. 

For comparison
with the intrinsic absorption column densities 
$N_{\rm H,intr}$ obtained by fitting the X-ray continuum, 
Table\,\ref{source_list} lists the column density $N$(HI) of the galaxy 
pair as measured in the 21-cm line VLA observations \citep{elmegreen95a}, 
averaged over the X-ray extraction aperture and not corrected
for helium. In addition to $N$(HI), there is a significant H$_2$ 
column density in much of this galaxy pair \citep{thomasson04}.
The discrete sources, X2, X3, X8, X9 have no significant intrinsic 
absorption along the line of sight, which places them in the layers of
the NGC 2207 gas disk closest to the observer. Source X1 may lie close to
the midplane of the NGC 2207 gas disk.
Most of the other discrete sources are more deeply embedded or located towards
the farther side of the NGC 2207 gas disk.

We have the following identifications or possible identifications of the
discrete X-ray sources (see Figure\,\ref{clumps} for the labelling of the
8 \micron\ or ultraviolet clumps): X5 is the nucleus of NGC 2207, X10 is
in the star-forming clump IR 11 on the eyelid of IC 2163, X1 coincides
with clump rc1, and X2 corresponds with clump IR 21, which contains SN 1999ec.
The value of $N_{\rm H,intr}$ for source X10 is consistent with
its location in clump IR 11 on the eyelid of IC 2163 behind an outer arm
of NGC 2207; the \HI\ column density of NGC 2207 at X10 is 
$0.32 \times 10^{22}$ atom cm$^{-2}$ and thus the fitted value of
$N_{\rm H,intr}$ places X10 behind NGC 2207 but
on the nearer side (relative to us) of the gas layer in the {\it eyelids}.
The coincidence between X10 and star-forming clump IR 11 suggests 
that X10 is a high mass X-ray binary. Alternatively, \citet{smith05}
note that some discrete X-ray sources in star forming regions may be due to
SNRs with high-mass progenitors, rather than high mass X-ray binaries.
Source X5 (the nucleus of NGC 2207) 
is the brightest X-ray source in the entire
field. As a matter of fact it is the only X-ray source in the hard X-ray band
above 5 keV. We shall discuss sources X5 and X1 in
Sections 6.2 and 6.3, respectively, and source X2 in Section 8.
 
The source denoted RR in Table\,\ref{source_list} is not a discrete
source but actually the entire
{\it NE radio ridge}, overlapping part of source X9. The  X-ray spectrum
of the {\it NE radio ridge} suggests that the emission could be mainly 
from a collection of X-ray binaries.    
Thus the {\it NE radio ridge} is not unusual either in terms of its
X-ray surface brightness or its X-ray spectrum. 
We do not detect SN 2003H on the
{\it NE radio ridge} as a discrete X-ray source.

\subsection{The nucleus of NGC 2207}

\begin{figure}
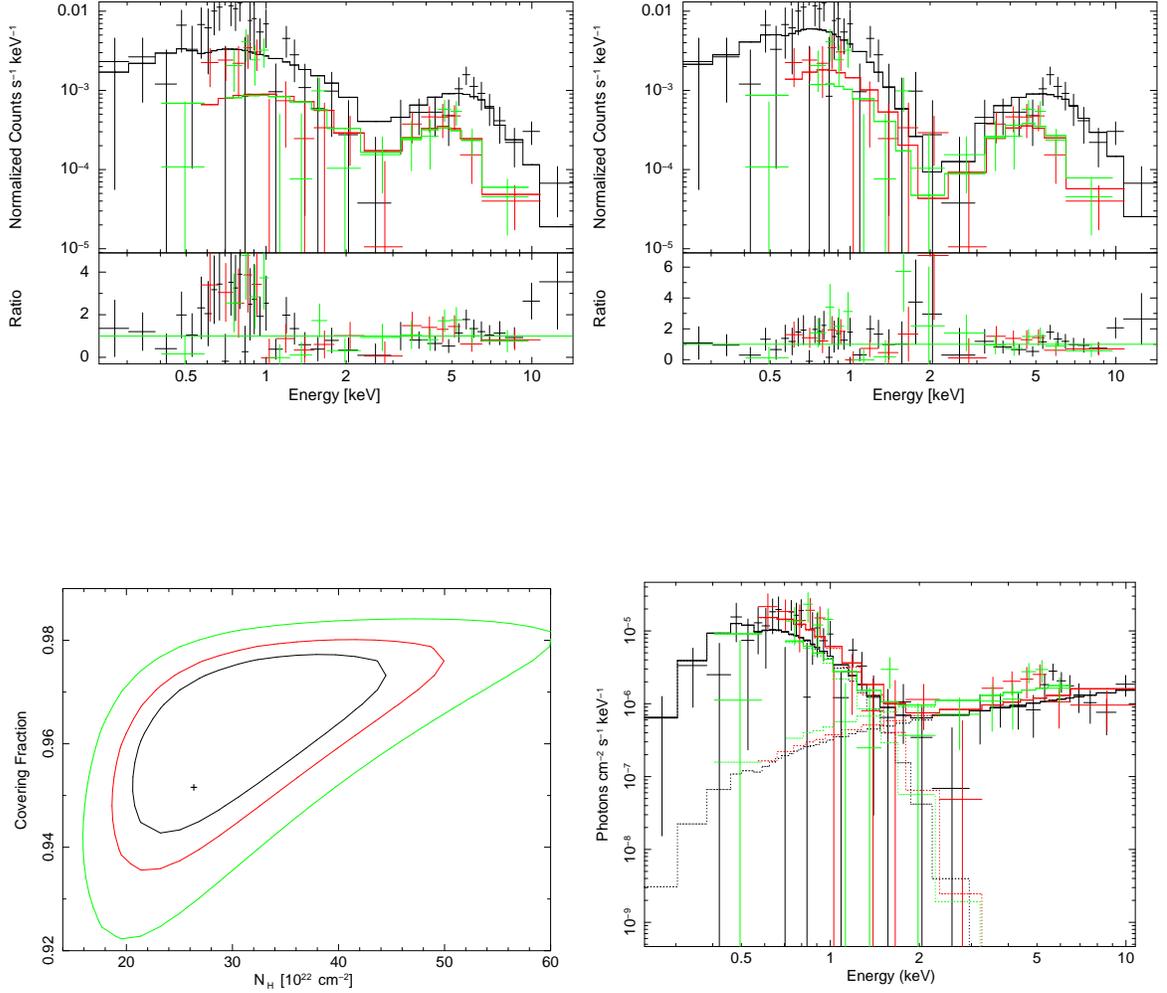

\epsscale{0.7}
\plottwo{xmm2207_f10a.ps}{xmm2207_f10b.ps}

\plottwo{xmm2207_f10c.ps}{xmm2207_f10d.ps}
\bigskip
\caption{\label{n2207_nucleus_xray_spec} EPIC pn (black) and MOS 1 and 2 (red
and green) spectra of the nucleus of NGC 2207 fitted by a power law model with
partial covering absorber (upper left panel) and blackbody plus absorbed power 
law spectrum (upper right panel). Note the hard X-ray emission at 5 -- 10 keV. 
Both fits include absorption by 
neutral gas in the Milky Way with an absorption column density of 
1.13$\times 10^{21}$ cm$^{-2}$ \citep
{dic90}. The lower left panel displays the contour plot for the absorption
 column density
and covering fraction of the partial-covering absorber model. The lower right
 panel shows the blackbody component and the power-law component separately for
the model in the upper right panel.
}
\end{figure}

Figure\,\ref{n2207_nucleus_xray_spec} displays the pn and MOS spectra of the
nucleus of NGC 2207 and shows that its emission contains a hard 
X-ray component. 
In order to increase the signal-to-noise ratio, all three
spectra were fitted simultaneously in XSPEC. For all fits the absorption column
density of the Milky Way was fixed to the Galactic value 
\citep[1.13$\times 10^{21}$ cm$^{-2}$; ][]{dic90}. Fitting a single power law
model to the data fails as the observed spectrum is double-humped; 
it yields a $\chi^2$/dof = 125/65 (where 
dof means degrees of freedom) and an  X-ray spectral slope 
that appears to be unphysically flat with \ax=--1.16.
The strong increase of the flux in hard X-rays
suggests that a strong partial-covering absorber is present in the 
NGC 2207 nucleus. Models involving partial covering of an AGN X-ray
source by dense blobs close to the central source
have often been used to interpret the spectra of 
Narrow-Line Seyfert 1 galaxies \citep[e.g.,][]{grupe07, turner09}.
We fitted a partial-covering absorber 
with a power law model to the data as shown in the upper left panel
of Figure\,\ref{n2207_nucleus_xray_spec}. For this model, the fit is
acceptable ($\chi^2/{\rm dof}$ = 83/65) and yields for the partial covering
absorber a column density 
of $N_{\rm H, pc}= 2.7\times 10^{23}$ cm$^{-2}$ and
covering fraction of $f_{\rm pc}$=0.95 as listed in Table\,\ref{x5_xspec}.
The lower left panel of Figure\,\ref{n2207_nucleus_xray_spec} displays the 
 contour plot between the column density and the covering fraction of the
absorber. From this model we derived an unabsorbed flux in the
observed 0.3-10.0 keV band of $2.1\times 10^{-13}$ ergs s$^{-1}$ cm$^{-2}$
which converts to a luminosity of
$0.4 \times 10^{41}$ ergs s$^{-1}$ in the 0.3-10.0 keV band. This is the
luminosity of a low-luminosity Seyfert galaxy, most-likely  Seyfert 2 galaxy.
There is excess emission at energies below 1
keV, which can be interpreted as strong emission lines from ionized gas. Such
emission lines from ionized gas have been reported in several cases of partial
covering absorption sources, such as the Narrow-Line Seyfert 1 
Galaxies Mkn 1239
and Mkn 335 \citep{grupe04,grupe07,grupe08} or the Seyfert 2 galaxy 
NGC 5643 \citep{guainazzi04}. Our data, however, do not allow 
us to constrain any
parameters of this ionized gas. 

The pn and MOS spectra can be fitted alternatively by an absorbed
 blackbody plus power law model with no intrinsic absorption. 
This fit is
shown in the upper right panel of Figure\,\ref{n2207_nucleus_xray_spec}
and listed in Table\,\ref{x5_xspec}. The lower right panel of
Figure\,\ref{n2207_nucleus_xray_spec} displays the blackbody and the 
power law components separately. The blackbody
temperature is $kT=163^{+29}_{-19}$eV and the energy spectral index
\ax=--1.56$^{+0.14}_{-0.16}$ with $\chi^2/{\rm dof}$=74/62.
Still, the residuals below 1 keV remain.
Although the $\chi^2/$dof of the blackbody plus power-law model 
suggests a slightly better fit compared with the partial-covering absorber 
model, 
an F-test shows this is only a slight improvement
(F- value = 2.51 and a probability of a null random result $P$= 0.067)
The energy spectral slope of the hard X-ray power law is unphysically
flat, i.e., would require too many higher energy relativistic
electrons compared to lower energy relativistic electrons. 
Therefore we conclude that the most likely model to explain the
X-ray spectrum of the nucleus of NGC 2207 is the power law model with 
partial-covering absorption.

The nucleus has an H$\alpha$ flux of $1.58\times 10^{-14}$
 ergs s$^{-1}$ cm$^{-2}$ and a \sixcm\ flux density of 0.37 mJy
(equivalent to a \sixcm\ radio continuum luminosity of
$5 \times 10^{19}$ W Hz$^{-1}$). These values are
similar to those of the {\it eyelid} clump IR 10. 
Half of the clumps in Table\,\ref{n2207clumps}
have greater values of the \sixcm\ flux density than the NGC 2207 nucleus,
and thus the nucleus is relatively radio quiet for an AGN. 

\subsection{Source X1}

\begin{figure}
\epsscale{0.9}
\plottwo{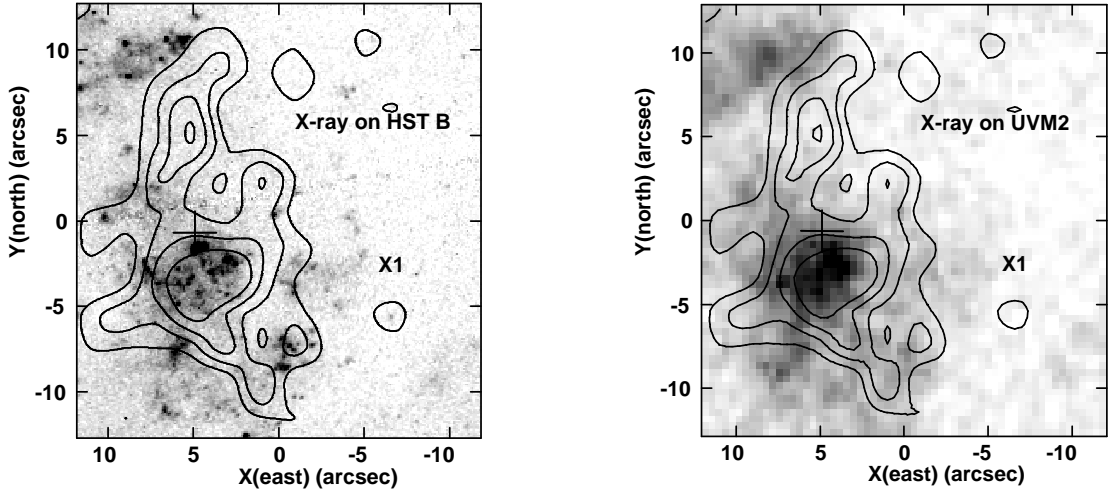}{xmm2207_f11b.ps}
\caption{\label{X1} For source X1, X-ray contours from the MOS data overlaid
on the {\it HST} $B$-band image (left panel) 
and on the $UVM2$ image (right panel).
The location of the unresolved radio clump rc1 is marked by 
a plus sign with width equal to the HPBW of the \sixcm\ synthesized beam. 
}
\end{figure}

Figure\,\ref{X1} provides a detailed view of X-ray source X1 
with X-ray contours from the MOS data 
overlaid on the {\it HST} $B$-band in the left panel and overlaid on the
$UVM2$ image in the right panel. The plus sign marks the location of 
the unresolved, nonthermal radio continuum clump rc1, and 
the width of the plus sign is the
HPBW of the \sixcm\ synthesized beam. The main X-ray emission is centered
on a collection of blue star clusters, the most prominent of which lies
close to the discrete radio source.
With $A_{\rm v}/N_{\rm gas}$ = 
$0.53 \times 10^{-21}$ mag per atom cm$^{-2}$
from \citet{bohlin78} for solar neighborhood metallicity, the intrinsic 
X-ray absorption column density of X1 is equivalent to
$A_{\rm v}$ = 0.6$^{+1.1}_{-0.6}$ mag.
The value of the 8 \micron\ to $UVM2$ flux density ratio 
(see Section 4) suggests that the clump rc1 suffers little extinction.
Other evidence of low extinction is the close resemblance between
the $B$-band and $UVM2$ images.
The \HI\ line profiles obtained by \citet{elmegreen95a} show 
no evidence for an absorption feature at rc1,
but the \HI\ data has low spatial resolution of $13.5'' \times 12''$.

Fitting a 2D Gaussian plus a flat baseline to the unresolved source rc1 
on our 
\sixcm\ radio continuum image with $2.5''$ resolution gives 
$S_\nu$(6 cm) = $1.12 \pm 0.03$ mJy.
If the radio source rc1 is in NGC 2207
(rather than a background quasar) and is isotropic, it has a 
\sixcm\ radio continuum luminosity $L_{\nu}$(6 cm) =
$1.6 \times 10^{20}$ W Hz$^{-1}$, which is $230 \times L_{\nu}$(6 cm) of
Cas A \citep[taking $L_{\nu}$(6 cm) of Cas A as $7 \times 10^{17}$ W Hz$^{-1}$
from][]{weiler89}. 
This \sixcm\ luminosity lies in the range of radio supernovae and in the
range of the very brightest SNRs.
 \citet{neff00} find the three brightest, discrete, nonthermal radio
sources in the Antennae merger pair have $L_{\nu}$(6 cm) in the range 
$8 \times 10^{19}$ to
$2 \times 10^{20}$ W Hz$^{-1}$ and attribute these to SNRs.
Each is slightly extended in their high resolution 
radio continuum observations. 
The brightest of these Antennae SNRs is listed as an ULX candidate 
by \citet{swartz04}
with a $L_{\rm x}$ in the 0.5 - 8 keV band of $18 \pm 11 \times 10^{39}$
erg s$^{-1}$ and $\Gamma$
= 2.54, i.e., similar in X-ray spectrum and somewhat more luminous in X-rays
than our source X1, which has $L_{\rm x}$ =$7 \times 10^{39}$ erg s$^{-1}$. 

Possible interpretations of the source
X1/rc1 are a radio supernova, a supernova remnant, a background quasar,
or a chance superposition of a ULX in NGC 2207 with a background radio
quasar. 
It is clear from Figure\, \ref{20cm-70micron} that rc1 
was brighter on 2003 Jan 14 when the $\lambda 3.5$ cm (8.46 GHz) radio
continuum observations were taken than on 1990 October 11 when the
line-free $\lambda 20$ cm radio continuum observations were made: in the 
$\lambda 3.5$ cm image, rc1 is much brighter than
every other source in the galaxy pair except {\it feature i}, whereas 
 no local surface brightness peak is seen at the 
location of rc1 in the $\lambda 20$ cm radio continuum image.  Also, no
local surface brightness peak at rc1 is visible in the figure in
\citet{condon83} which displays a $\lambda 20$ cm radio continuum image of
NGC 2207/IC 2163 with $7.6'' \times 6''$ resolution. For a quantitative 
comparison, we measured the flux density of a
$21'' \times 21''$ box centered on rc1 in the above $\lambda 3.5$ cm image
and in a \sixcm\ image made from the uv-data used in 
Figure\, \ref{n2207_6cm}, i.e., B-configuration (high resolution) observations
on 2001 April 14 plus D-configuration (low resolution) observations on
1995 May 13. For the $21''$ box,  $S_\nu$(6 cm) = 3.38 mJy and  
$S_\nu$(3.5 cm) = 3.26 mJy. Part of this flux density is from emission 
unrelated to rc1. The \sixcm\ flux density unrelated to rc1 in this box 
equals 3.38 mJy -- 1.12 mJy = 2.26 mJy ( where $S_\nu$(6 cm) =
1.12 mJy for rc1, as measured on the high resolution image).
 If the spectral index $\alpha$ 
of the unrelated emission lies in the range 0.1 to 0.9, then scaling the
2.26 mJy from \sixcm\ to  $\lambda 3.5$ cm gives $S_\nu$(3.5 cm) in the
range 2.14 to 1.37 mJy for the unrelated emission. Subtracting this from the
measured $S_\nu$(3.5 cm) = 3.26 mJy of the $21''$ box in 2003 gives 
 $S_\nu$(3.5 cm) for rc1 in 2003 in the range 1.12 mJy to 1.69 mJy.
As this is greater than or equal to the \sixcm\ flux density of rc1 in the
high resolution observations in 2001, either rc1 had increased in
radio luminosity by 2003 and/or there is synchrotron self-absorption in
the radio. Thus it is unlikely that rc1 is an SNR. The source rc1 
could be a background quasar or a radio supernova. 
A young radio supernova whose spectrum is inverted because the
emission at lower frequencies is still absorbed 
\citep{weiler88} is an example of a radio 
source with synchroton self-absorption that would be consistent with 
the observed properties of rc1.

The X-ray, radio, and ultraviolet emission from X1/rc1 may arise from
parts of the source that differ in column density. For example 
\citet{bregman92} find that the X-ray and radio observations of the very
luminous SN 1986J in NGC 891 are consistent with a model in which the
radio emission is from the high temperature outward-moving shock whereas
the soft X-ray emission is from the cooled material associated with the
reverse shock.  

\section{\label{N2276} Comparison between NGC 2207 and NGC 2276}

We compare NGC 2207 with the spiral galaxy NGC 2276 since both galaxies 
are unusual in having
a long ridge of bright radio emission from the outer part of the disk
on one side of the galaxy.
In NGC 2207 we call this feature the {\it NE radio 
ridge}. The NGC 2276 radio continuum ridge is prominent in the images
displayed in \citet{hummel95} and \citet{condon83}.
Globally the NGC 2207/IC 2163 pair and NGC 2276 each have a radio
continuum flux density that is 2.5 to 3 times higher than expected from
the IRAS far-infrared flux. 

NGC 2276 is a member of a small group
of galaxies embedded in a large diffuse, intragroup, X-ray cloud.
Unlike the {\it NE radio ridge} in NGC 2207, the large-scale
bow-shock-like radio continuum ridge along the western edge of NGC 2276 is
a site of active star formation and bright X-ray emission, visible in 
the \chandra\ observations by 
\citet{rasmussen06} and the ROSAT High-Resolution Image by 
\citet{davis97}. The stellar and gaseous disks
in NGC 2276 truncate just beyond this shock-like feature whereas the
{\it NE radio ridge} in NGC 2207 is not the outermost spiral arm on the
affected side of NGC 2207. The lopsided appearance of NGC 2276 has been 
attributed either to a tidal interaction with the ellipical galaxy NGC 2300
\citep{hummel95,davis97} or to ram pressure from the hot intragroup
gas \citep{rasmussen06}. From \chandra\ X-ray observations,
\citet{rasmussen06} find that NGC 2276 is moving supersonically at
850 \kms\ through the hot intragroup gas and that the ram pressure,
which may have been acting for several $\times 10^8$ yr, could explain
the observed compression of the \HI\ gas and magnetic fields along the
western edge of NGC 2276, leading to active star formation in this
region. They argue that the X-ray emission from the western edge of NGC 2276
is dominated by hot plasma resulting from the vigorous star formation.
This may be the reason why the {\it NE radio ridge} in NGC 2207 is not bright
in X-rays, since it is not a site of extended vigorous star formation.
If the scraping between NGC 2207 and IC 2163 and the bright radio continuum 
emission
from the {\it NE radio ridge} are high off the disk of NGC 2207, this would
explain why the observed \HI\ column density
contours are not unusually compressed on
the {\it NE radio ridge} and why there is no widespread active star formation
there.

The IC 2163/NGC 2207 encounter models in \citet{struck05} 
predict a gas stream from IC 2163
impinging on NGC 2207, with shocks at a few hundred \kms\
being pushed into both galaxies.
However,  the magnitude and direction of 
the tidal forces change as the encounter proceeds, and thus the 
compressed magnetic field in the {\it NE radio ridge} may be a 
short-lived transient phenomenon.

\section{\label{feature i} Feature $i$}

\begin{figure}
\epsscale{1.0}
\plottwo{xmm2207_f12a.ps}{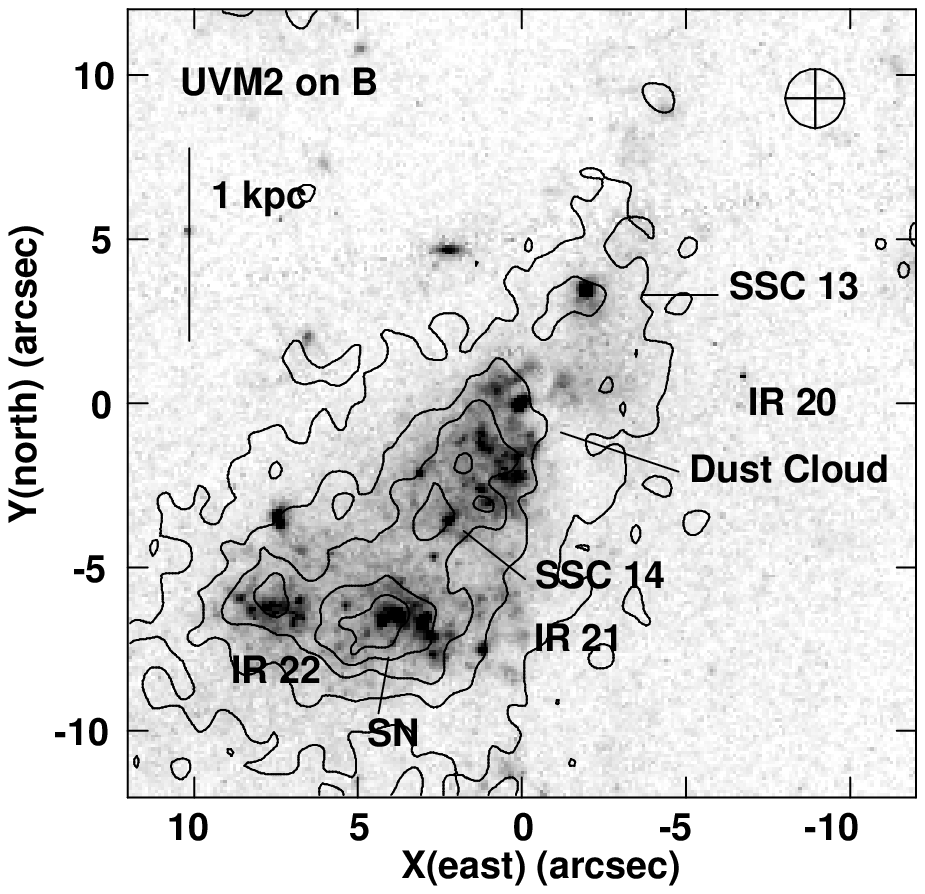}
\plottwo{xmm2207_f12c.ps}{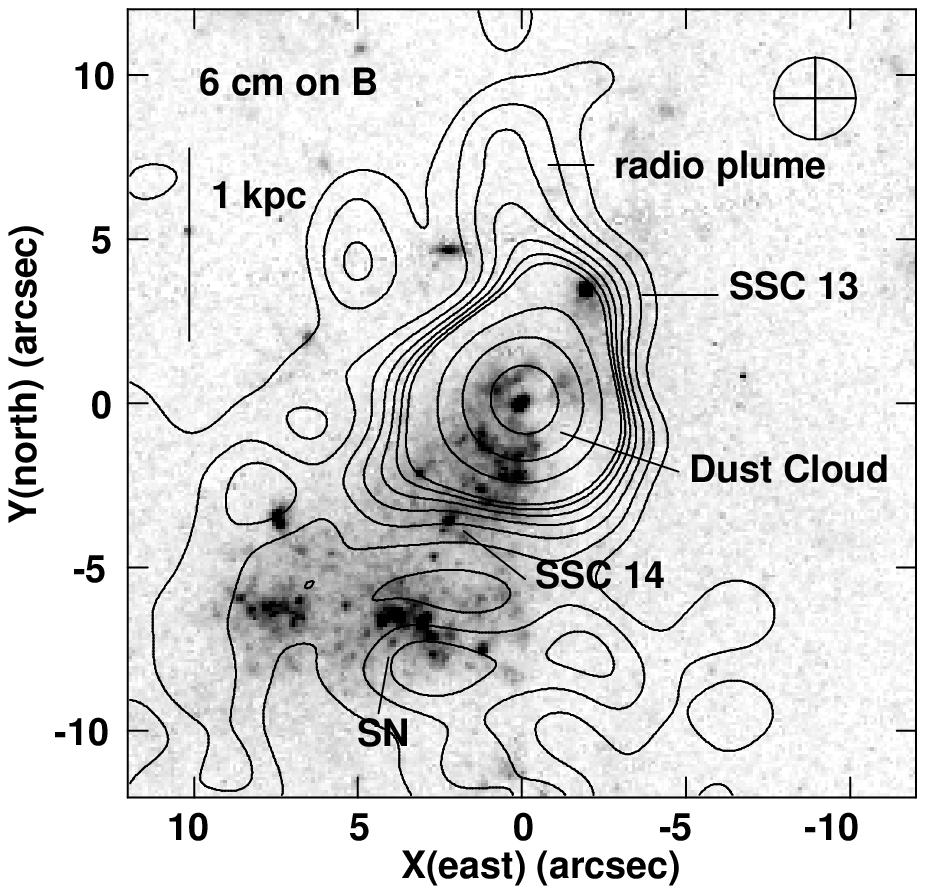}
\caption{\label{featuri_images} Overlays on the {\it HST} $B$-image of 
{\it feature i} at
different wavelengths. The upper left panel displays the contours of the X-ray
emission, the upper right the {\it UVM2} data from the \xmm\ OM observations, 
the lower left 
the 8 \micron\ {\it Spitzer} observations, and the lower right the 
\sixcm\ radio continuum
emission. The 8 \micron\ contours are at 2...(2)...10...(5)...40, 60, 80 and
100 MJy sr$^{-1}$. The \sixcm\ radio contours are at 4...(2)...16, 32, 64, 128
times the rms noise of 0.016 \mJybeam. The label
SN points to a stellar image at the SW edge of the $r$ = 0.3\arcsec\
error circle that \citet{vandyk03} give for the position of SN 1999ec;
they conclude that this star is too bright to be the progenitor.
}
\end{figure}

{\it Feature i} is the most luminous radio continuum, 
8 \micron, 24 \micron, and
H$\alpha$ source in the galaxy pair.
Figure\,\ref{featuri_images} displays the contours of emission from
{\it feature i} and environs in various wavebands overlaid on the {\it HST}
$B$-band image from observations made in 1996. 
As shown in this figure, {\it feature i} is bright in
\sixcm\ radio continuum and 8 \micron\ emission but highly absorbed
in the $UVM2$ band. Comparison of the $UVM2$ and 8 \micron\
images of this region provides an extinction map. 
\citet{elmegreen00} point out the 
opaque (optically thick even in $I$-band) conically-shaped dust cloud
(labelled here) with a bright compact cluster at its apex. The core
of the radio source and the brightest 8 \micron\ and 24 \micron\
emission are centered on this cluster. The $UVM2$ emission from the radio core
and the conical dust cloud is highly absorbed and no X-ray
emission is detected from either 
(see Figure\,\ref{featuri_images}).

In the radio continuum, {\it feature i} is a core plus envelope source.
Aside from a northern plume in the radio, it looks very similar
in the \sixcm\ and 8 \micron\ images: the extended emission
fills a triangular region with E-W base $8''$
and N-S height $9.6''$ (= 1.4 kpc $\times$ 1.7 kpc), which includes the
cluster arcs of \citet{elmegreen00} and two super-star clusters
identified by \citet{elmegreen01}. Just north of the filled triangular region,
the \sixcm\ emission forms a plume with position angle PA = 5\arcdeg, 
whereas the 8 \micron\ emission is a little west of north, i.e., along the
arm at PA = -15\arcdeg. We take as the definition of {\it feature i} the
filled triangular region plus the radio plume. It has $S_\nu$(6 cm) =
4.67 mJy

We fit a simplified model consisting of the sum of 
two 2D Gaussians plus a flat baseline
to the \sixcm\ emission from {\it feature i} to represent the core plus
envelope plus general arm emission. Table\,\ref{featuri_model} lists the
results obtained by using our \sixcm\ image with 
the circular 2.5\arcsec\ synthesized
beam and the results obtained by using
our original \sixcm\ image which has higher resolution in the E-W
direction (synthesized beam = $2.48'' \times 1.30''$, BPA = 8\arcdeg).
 These Gaussian models give
for the core plus envelope $S_{\nu}$(6 cm) =$4.6 \pm 0.07$ mJy, attribute
roughly 60\% of the emission to the core, and find the core is
slightly elongated along the same line as the opaque dust cloud
(position angles 40\arcdeg\ for the core and about 
40\arcdeg\ + 180\arcdeg\ for the dust cloud). The opaque dust
cloud has a projected length of 
2\arcsec\ -- 3\arcsec (about twice the diameter
of the radio core). 

From VLA B configuration snapshot observations in 1986,
\citet{vila90} measured $S_{\nu}$(6 cm) = 1.4 mJy for the $1''$ core
and 3.4 mJy for {\it feature i} as a whole, whereas our combined 
VLA \sixcm\ data from B configuration (high resolution) 
observations in 2001 and 
D configuration (low resolution) observations in 1995 give 2.66 mJy 
for the $1''$ core and 4.66 mJy for {\it feature i} as a whole
on our highest resolution image (see Table\,\ref{featuri_model}), i.e.,
1.3 mJy greater than the Vila et al. value for the core and 1.3 mJy
greater than the Vila et al. value for {\it feature i} as a whole. 
Differences in flux density can arise from the missing short spacings and 
higher noise in the Vila et al. data and different ways of measuring 
the source. However since the difference in flux
density between 1986 and 2001 
is the same for the core and for {\it feature i} as a whole and
the \sixcm\ flux density of the core in 2001 is nearly twice the
Vila et al. value, we
conclude that the \sixcm\ flux density of the core increased by 1.3 mJy
between 1986 and 2001. This corresponds to an increase in $L_{\nu}$(6 cm) of
$1.9 \times 10^{20}$ W Hz$^{-1}$ if isotropic.
As this lies in the luminosity range of radio supernovae, it seems likely
that in 2001 a radio supernova was present in the core. From these data,
we cannot determine the year when outburst in the core of {\it feature i}
occurred; \citet{weiler88}
point out examples of radio supernovae that remained bright in
the radio for a  number of years. A supernova in the core of {\it feature i} 
may have been hidden from view optically by the high extinction. If
our interpretation is correct, 
 then NGC 2207 is remarkable in
having had two optical supernovae (SN 1999ec and SN 2003H, both Type Ib)
plus one radio supernova in recent years (i.e., between 1986 and 2003)
 and all with high-mass progenitors.

 The simulation models by \citet{elmegreen95b} and
\citet{struck05} for the NGC 2207/IC 2163 encounter estimate 
the inclination $i$ of the main disk of NGC 2207 as 25\arcdeg\ -- 35\arcdeg\
(relative to face-on) with the minor axis of the projection at 
PA = 50\arcdeg\ -- 70\arcdeg. The near side (relative to us) of NGC 2207 
is the northeastern side. The  opaque dust cloud is aligned nearly
parallel to the minor axis of the projection of the main disk of NGC 2207
into the sky-plane.
\citet{elmegreen00} point out a red V-shaped structure with 
apex at the radio core and opening
to the north (see Fig. 6 of that paper).
The left fork of the V appears to be a continuation of the opaque dust
cloud to the opposite side of the core, whereas the right fork of
the V is aligned with the inside edge of the optical arm farther north,
but straighter than most spiral-arm dust lanes, possibly
as a result of extra compression of the 
original dust
lane by the energetic events in {\it feature i}. 
The radio plume is midway between the two forks of the V. 
Given the orientation of the main disk of NGC 2207, the opaque dust cloud 
plus the left fork of the V could be outflow perpendicular to the plane of
NGC 2207, generated at the radio core, i.e., the dark dust cone could be gas 
approaching us on the near side (relative to us) of the midplane and
the left fork of the V could be gas receding from us on the far side and thus
less prominent as an absorption feature because not obscuring light on
the near side of the midplane.

\citet{elmegreen06} measure a 24 \micron\ flux density $S_\nu$(24 \micron) =
248 mJy for the $24'' \times 24''$ field displayed in 
Figure\,\ref{featuri_images}. In the HiRES 24 \micron\ image in
\citet{velusamy08} (which has a resolution of 1.9\arcsec, a little
better than that of the 8 \micron\ and \sixcm\ images shown here),
this flux density comes from a $7.5'' \times 7.5''$
region centered on the radio core of {\it feature i}. This gives
a 24 \micron\ to \sixcm\ flux density ratio for 
{\it feature i} of 248 mJy/4.67 mJy = 53,
which is somewhat greater than the mean value 
$S_\nu$(24 \micron)/$S_\nu$(6 cm) = $36 \pm 9$
for the M81 \HII\ regions
in Table\,\ref{m81} but similar to that of the most luminous giant radio 
\HII\ region (K181) in M81. In Section 4
(see also Figure\,\ref{n2207_8m_6cm}) we found that {\it feature i} is
underluminous at 8 \micron\ relative to its \sixcm\ radio continuum emission
when compared with the mean value for the  M81 \HII\ regions and with 
the mean value for clumps in 
NGC 2207 containing OB associations. For {\it feature i},
$S_\nu$(24 \micron)/$S_\nu$(8 \micron) = 248 mJy/35 mJy = 7.1, which is
high compared to the mean value of 1.9 for the M81 \HII\ regions in
Table\, \ref{m81} and compared to the mean value for the SINGS sample
(see Table\, \ref{q}), but similar to that of the dwarf (Im) starburst galaxy
Mrk 33, which has $S_\nu$(24 \micron)/$S_\nu$(8 \micron) = 6.6 \citep{dale07}.
Emission at 24 \micron\ is from warm
very small grains whereas emission at 8 \micron\ is a combination of PAH 
emission bands
and continuum emission from warm very small grains. The
high value of the 24 \micron\ to 8 \micron\ flux density ratio of 
{\it feature i} and the
low value of the 8 \micron\ to \sixcm\ flux density ratio suggest some  PAH 
destruction by the radiation field of {\it feature i} 
has depressed its 8 \micron\ emission. 

In the HiRES 24 \micron\ image in \citet{velusamy08}, 
{\it feature i} appears as a filled elliptically-shaped region with
minor axis/major axis ratio = 0.8 and minor axis at PA = 50\degr.  
Given the  orientation of the disk of NGC 2207, it is probably a filled
circular region in the disk of NGC 2207.  

A long-slit optical spectrum taken by Pierre Martin (private communication)
with the CFHT MOS prior to 2000 cuts E-W through the core of
{\it feature i} and exhibits a normal \HII\ region spectrum with no unusual
line ratios. The long-slit optical spectrum of 
SN 1999ec (8\arcsec\ SSE of the 
radio core) taken by \citet{matheson01} has a PA of -17\arcdeg,
a resolution of 6.3 \AA\ (380 \kms) FWHM at $\lambda 5000$ \AA,
includes cluster arcs in {\it feature i}, and crosses the east
fork of the red V. North of SN 1999ec, this spectrum shows normal \HII\
region emission but the line profiles have a little asymmetry with a
slightly more extended red wing, i.e., for the [\ion{O}{3}]
$\lambda 5007$ \AA\ line, the center of a Gaussian fit is shifted 
systematically to the red relative to the intensity maximum with the
shift increasing from $0.2 \pm 0.3$ \AA\ at the southern cluster arcs to
$0.6 \pm 0.3$ \AA\ at the east fork of the red V. This may be an 
instrumental effect associated with non-centered sources projecting onto
different positions on the chip. The optical emission-line velocities
are generally consistent with the \HI\ velocities given that the \HI\
data from \citet{elmegreen95a} has low spatial resolution and a velocity
dispersion of 56 \kms\ at {\it feature i}. Neither of these optical
spectra is suitable for looking for low velocity outflows from 
{\it feature i}. At the {\it feature i} positions sampled by these
two long slits, photoionization dominates, and these optical spectra
show no components in {\it feature i} at supernova or jet velocities. 

Since photoionization dominates the optical spectra whereas the radio emission
is nonthermal, we interpret {\it feature i} as a mini-starburst and
compare it with the central starburst in M82.
 The central $50'' \times 15''$ starburst in M82 has
$S_\nu$(6 cm) equal to $3.4 \pm 0.2$ Jy \citep{hargrave74}, a radio spectral 
index $\alpha$ of 0.5, and outflow over a wide range of solid angle
\citep{seaquist91}. If the M82 central starburst were at the 35 Mpc 
distance of NGC 2207 instead of 3.6 Mpc, it would have a major axis of
$5.1''$ (a little smaller than {\it feature i}) and 
$S_\nu$(6 cm) = 36 mJy, which is 7.7 times the \sixcm\ flux density of
{\it feature i} as a whole. Except in the 1\arcsec\ core of {\it feature i},
the \sixcm\ radiation field in {\it feature i} is significantly less 
intense than the average value for the M82 starburst.
We estimate the value of 
$S_\nu$(24 \micron)/$S_\nu$(6 cm) of the central starburst in M82 for
comparison with {\it feature i}.
 With a 25\arcsec\ aperture centered on the M82 starburst, 
\citet{kleinman70} measured $S_\nu$(22 \micron) = 120 Jy. Extrapolating to
24 \micron\ by taking $I_\nu \propto \nu^{-1}$ gives 
$S_\nu$(24 \micron) = 130 Jy for a 25\arcsec\ aperture. For the
50\arcsec\ extent of the radio emission from the M82 starburst, 
$S_\nu$(24 \micron) should be somewhat greater than 130 Jy, and thus
$S_\nu$(24 \micron)/$S_\nu$(6 cm) should be somewhat greater
than 130 Jy/3.4 Jy = 38. It appears that the ratio
$S_\nu$(24 \micron)/$S_\nu$(6 cm) = 53 for {\it feature i} is not unusual for 
a dusty starburst region.

The upper left panel in Figure\,\ref{featuri_images}
 displays contours of the X-ray emission from source X2. The core of
{\it feature i} is not detected as an X-ray source. The 
absence of soft X-rays could be the
result of the high absorption seen from comparison of the $UVM2$ and
8 \micron\ images. The X-ray emission is generally south and SSE of  
{\it feature i}, and the brightest X-ray knot is centered 2.4\arcsec\
north of SN 1999ec. The upper right panel in this figure shows that the
$UVM2$ peak emission coincides with the supernova. 
As previously reported by \citet{pooley07} from an
archival search of Type Ib,c supernovae which used our \xmm\ observations
of NGC 2207, it seems likely that most of the X-ray emission from X2 is
associated with the Type Ib supernova SN 1999ec. It is unusual for a 
Type Ib supernova to be seen as an X-ray source, and, interestingly, 
SN 1999ec is bright in X-rays 6 years after the optical SN was discovered.
The X-ray emission may be from the supernova shock plowing into a
circumstellar envelope. This makes it more plausible that the 
bright $UVM2$ emission coinciding with SN 1999ec is due to shock excitation
of circumstellar gas by the supernova. 
 The X-ray spectral index of X2 is steeper than typical of
an X-ray binary.

\section{\label{discuss} CONCLUSIONS}

We presented the X-ray and UV data observed by \xmm\ and new \sixcm\ radio 
continuum observations of the interacting galaxies NGC 2207/IC 2163.
When combined with our previous observations in H$\alpha$, {\it HST} $B$-band,
{\it Spitzer} infrared, \HI, and {\it SEST} \CO, these data allow us to
see the effects of the grazing encounter in producing large scale shocks,
to study star complexes, supernovae, and the galactic nuclei in this pair,
and to identify possible ULX candidates.

In X-rays we detect the nucleus of NGC 2207, nine possible ULX candidates,
 and extended X-ray emission,
mainly from NGC 2207. One of the discrete X-ray sources corresponds to
SN 1999ec and another has brightened in the radio in recent years and
could be a
radio supernova or a background quasar. The bright $UVM2$ and X-ray emission 
from SN 1999ec may be from shock excitation of circumstellar gas by the
supernova.
The strongest source in our \xmm\ X-ray observations of NGC 2207/ IC 2163 is 
the nucleus of NGC 2207. It is the only hard X-ray source in the field. 
The preferred model for its X-ray spectrum  is a power law with a
partial covering absorber. 
Most likely this is a strongly absorbed, low-luminosity, Seyfert 2 AGN. 
However, optical
spectroscopy is needed to confirm this assumption.

We measured values of the ratio of 8 \micron\ to \sixcm\ radio continuum
flux density for the prominent, kpc-sized, star-forming clumps in the galaxy 
pair and compared them with those for the M81 \HII\ regions whose 
radio continuum 
is dominated by optically-thin free-free emission. For the
bright clumps in NGC 2207, the mean value of this ratio equals 18, which is
the same as for giant radio \HII\ regions in M81, within the
uncertainties.  For the bright clumps
on the rim (the {\it eyelids}) of the eye-shaped oval in IC 2163, 
the mean value is nearly a factor of two greater. 

There are two types of large-scale fronts in this galaxy pair: the 
{\it eyelids} of the eye-shaped oval in IC 2163 and the 
{\it NE radio ridge} on the companion side of NGC 2207. The {\it eyelid}
shock front is produced by radial inflow and convergence of orbits and has 
vigorous star formation in a dusty environment. In the {\it eyelids} the ratio
of 8 \micron\ to \sixcm\ surface brightness is appreciably greater than in
the M81 giant \HII\ regions. This could be the result of heating mainly by
$B$ stars instead of $O$ stars and/or fragmentation of dust
grains by collisions in the shock region and formation of more PAH
molecules close to the $OB$ stars. 
The {\it eyelids} are located behind the outer part of NGC 2207; this may
explain why they are not bright in extended soft X-ray emission.
 
The {\it NE radio ridge} in NGC 2207 is particularly
bright in the radio continuum but not in any of the tracers of recent
star formation. Values of the 
ratios of 8 \micron\ to \sixcm\ surface brightness and 24 \micron\ to
\sixcm\ surface brightness are low on the {\it NE radio ridge} compared to
those of the M81 \HII\ regions. 
Unlike the bright radio ridge in the outer part of NGC 2276, the 
{\it NE radio ridge} is not enhanced in extended X-ray emission
relative to the rest of galaxy; this is
probably because it is not a site of active star formation.
The {\it NE radio ridge}, which previous models attributed to disk
or halo scraping, is
simply due to compression of the  magnetic field and may be mainly
in the halo on the back side of NGC 2207 (between the two galaxies).
If the bright radio continuum emission from the {\it NE radio ridge} 
originates high off the disk of NGC 2207 in a region with little neutral gas, 
that would explain the lack of widespread vigorous star formation and the lack
of unusually compressed \HI\ column density contours 
on the {\it NE radio ridge}.   

For NGC 2207/IC 2163, the global values of the ratios of infrared to radio 
continuum flux density in the {\it Spitzer} 8 \micron, 24 \micron, and
70 \micron\ bands and the IRAS FIR are significantly below the medians/means
for large samples of galaxies. This is the result of excess radio continuum
emission from large portions of NGC 2207, not just the {\it NE radio ridge}.  

We find evidence that a radio supernova was present in the core of
{\it feature i} in 2001. If so, then NGC 2207 had two optical supernovae and
one radio supernova in recent years. The \sixcm\ radio
continuum luminosity of {\it feature i} on an outer arm of NGC 2207 is 13\%
of the central starburst in M82. In linear size, {\it feature i} is a little
larger than the M82 starburst. Like the M82 starburst, {\it feature i} is
a dusty starburst region with radio continuum emission that is mainly
nonthermal and with a 24 \micron\ to \sixcm\ flux density ratio somewhat
greater than the mean value for the M81 giant \HII\ regions. Whereas the M82 
starburst has outflow perpendicular to the disk which is bright in
X-rays, our only indication of outflow perpendicular to the disk in
{\it feature i} is the peculiar morphology of the dust structures.

\medskip
\acknowledgments
This work is based on observations obtained with \xmm, an ESA
 science mission with  instruments and contributions
 directly funded by ESA Member States and NASA.
We gratefully acknowledge support from NASA Goddard Grant 
NNG05GR10G to M.K., D.G., and D.M.E.
This research has made use of the NASA/IPAC Extragalactic
Database (NED) which is operated by the Jet Propulsion Laboratory,
Caltech, under contract with the National Aeronautics and Space
Administration. We also used a radio continuum image from the NRAO VLA
public archives. 
We thank Tom Matheson and Perry Berlind for sending us
FITS images of their long-slit spectrum of SN 1999ec. We thank Thangasamy
Velusamy for sending us the HiRes FITS image of the {\it Spitzer} 
24 \micron\ emission. We thank Pierre Martin for providing figures
displaying his long-slit spectrum of the core of {\it feature i} and 
his measurements of line ratios.

\clearpage

\begin{deluxetable}{llcccr}
\tabletypesize{\small}
\tablecaption{\xmm\ Observations of NGC 2207/IC 2163 on 2005 August 31
\label{obs_log}}
\tablewidth{0pt}
\tablehead{
\colhead{Instrument} & \colhead{mode/filter} &
\colhead{T-start\tablenotemark{a}} & 
\colhead{T-stop\tablenotemark{a}} &
\colhead{$\rm T_{exp}$\tablenotemark{b}} 
} 
\startdata
EPIC pn    & Ext. FF & 07:14:11 & 20:36:01 & 47640 \\
EPIC MOS-1 & FF      & 06:13:14 & 20:25:41 & 51567 \\
EPIC MOS-2 & FF      & 06:13:14 & 20:36:01 & 51572 \\
OM         & U       & 06:17:52 & 07:37:32 &  4460 \\
OM         & U       & 07:37:38 & 08:57:28 &  4479 \\
OM         & B       & 08:57:24 & 10:23:15 &  4840 \\
OM         & B       & 10:23:16 & 11:43:01 &  4460 \\
OM         & UVW1    & 11:43:02 & 13:02:47 &  4460 \\
OM         & UVW1    & 13:02:48 & 14:22:23 &  4460 \\
OM         & UVM2    & 14:22:34 & 15:42:19 &  4460 \\
OM         & UVM2    & 15:42:20 & 17:02:05 &  4460 \\
OM         & UVM2    & 17:02:06 & 18:21:51 &  4479 \\
\enddata

\tablenotetext{a}{Start and End times are given in UT}
\tablenotetext{b}{Observing time given in s}
\end{deluxetable}

\begin{deluxetable} {lcl}
\tabletypesize{\footnotesize}
%\tablenum{2}
\tablewidth{0pt}
\tablecaption{Images of NGC 2207/IC 2163
\label{psf}}
\tablehead{
\colhead{Image}  &\colhead{FWHM of PSF}  &\colhead{Reference}
}
\startdata
{\it HST} $B$ WFPC2  & $\sim 0.18''$\tablenotemark{a} &  \citet{elmegreen01}\\
$UVM2$          & $1.8''$                        &  this paper\\
$UVW1$          & $2.0''$                        &  this paper\\
MOS X-ray       & $\sim 5''$                     &  this paper\\
MOS + PN X-ray  & $\sim 9''$                     &  this paper\\
{\it Spitzer} 8 \micron\    & $2.4''$            &  \citet{elmegreen06}\\
{\it Spitzer} 24 \micron\   & $6''$              &  \citet{elmegreen06}\\
{\it Spitzer} 70 \micron\   & $18''$             &  \citet{elmegreen06}\\
{\it Spitzer} 160 \micron   & $40''$             &  \citet{elmegreen06}\\
$\lambda 3.5$ cm radio continuum & $11.7'' \times 3.5''$ & VLA public 
archives\\
\sixcm\ radio continuum & $2.48'' \times 1.3''$  &  this paper\\ 
\sixcm\ radio continuum    & $2.5''$             &  this paper\\
\sixcm\ radio continuum    & $6''$               &  this paper\\
$\lambda 20$ cm radio continuum & $13.5'' \times 12''$ & \citet{elmegreen95a}\\
H I                             & $13.5'' \times 12''$ & \citet{elmegreen95a}\\
SEST \CO\                  & $43''$              & \citet{thomasson04}\\
H$\alpha$                  & $4.2'' \times 3.6''$ & \citet{elmegreen01}\\
\enddata

\tablenotetext{a}{$\sim 0.18''$ for wide field and $\sim 0.09''$ for planetary
camera}
\end{deluxetable}

\begin{deluxetable} {lcccccccc}
\tabletypesize{\footnotesize}
%\tablenum{3}
\tablewidth{0pt}
\tablecaption{Spitzer and Ultraviolet Clumps in NGC 2207/IC 2163
\label{n2207clumps}}
\tablehead{
\colhead{Clump\tablenotemark{a}} 
&\colhead{$S_\nu$(6 cm)}
&\colhead{$S_\nu$(8 $\mu$m)} 
&\colhead{(8 $\mu$m/6 cm)\tablenotemark{b}} 
&\colhead{$S$(H$\alpha$)\tablenotemark{c}}
&\colhead{$A_v$\tablenotemark{d}}
&\colhead{$UVM2$}
&\colhead{(8 $\mu$m/$UVM2$)\tablenotemark{e}}
&\colhead{$UVM2-UVW1$}\\
& \colhead{(mJy)} & \colhead{(mJy)}
& &
& \colhead{(mag)}
& \colhead{(mag)} &  & \colhead{(mag)}
}
\startdata
IR 1  &  0.51&  10.4&  21&  0.59&  6.1&  18.6&  370 &  0.49\\
IR 2  &  0.36& 11.9&   33&  0.24&  6.9& $>$ 19.2& $>$ 710& $>$ 0.00\\
IR 3  &  0.63& 13.5&   21&  0.69&  6.2&  19.0&  650 &  0.58\\
IR 4  &  0.39& 12.5&   32&  2.6&   3.7&  17.7&  190 &  0.38\\
IR 5  &  0.11& 8.2&    72&  4.4&   1.2&  16.9&   62 &  0.10 \\
IR 8  &  0.22& 10.3&   47&  0.82&  4.4&  16.6&   59 &  0.06\\
IR 9  &  0.75& 19.3&   26&  2.4&   4.7&  16.7&  120 &  0.17\\
IR 10 &  0.34& 15.0&   44&  1.5&   4.3&  17.4&  170 &  0.24\\
IR 11 &  0.46& 11.2&   25&  2.0&   4.3&  18.0&  220 &  0.13\\
IR 81 &  0.41& 9.6&    23&  1.9&   4.1&  18.4&  280 &  $-0.23$\\
IR 6  &  0.10& 5.5&    57& \nodata & \nodata & $>$ 19.2&  $>$ 330 & $>$ 0.59\\
IR 13 &  0.34& 5.0&    15&  1.7&   4.1&  16.8&  34  &  0.03\\
IR 14 &  0.42& 6.9&    17&  2.7&   3.7&  16.7&  42  &  0.10\\
IR 15 &  0.41& 4.7&    11&  0.88&  5.2&  17.4&  56  &  0.06\\
IR 16 &  0.65& 9.4&    14&  2.9&   4.2&  17.3&  100 &  0.21\\
IR 17 &  0.56& 6.2&    11&  3.9&   3.6&  16.9&  43  &  0.01\\
IR 18 &  0.50& 9.8&    20&  3.4&   3.6&  17.3&  100 &  0.21\\
IR 19 &  0.53& 11.5&   21&  4.7&   3.3&  16.4&  53  & $-0.01$\\
IR 12\tablenotemark{f} &  0.81& 6.1&    7.3& 1.1&   5.9&  17.8&  100 & 0.35\\
IR 20\tablenotemark{g} &  4.39& 35.1& 8.0& 11.6&  4.9& 16.8&  240 & 0.23\\
IR 21\tablenotemark{h} &  0.28& 5.3&  19&  1.9&   3.6& 16.3&  22 & $-0.09$\\
IR 26 &  0.27& 5.7&     21&  2.1&   3.5& 16.5&  28 &  $-0.07$\\
IR 30 &  0.25& 5.9&   24&  1.8&   3.5&   17.8& 95 & 0.14\\
IR 32 &  0.23& 4.8&   21& \nodata & \nodata & $>$ 19.2& $>$ 290 & $>$ 0.14\\
IR 114&  0.34& 5.7&   17&  3.2&   3.2&   17.0&  43 & $-0.01$\\
u1    &  0.13& 5.6&   42& \nodata & \nodata &   16.2&  22 & $-0.03$\\
IR 138&  $<$ 0.04 & 0.81& \nodata & \nodata & \nodata & 17.0& 6.2 & $-0.25$\\
rc1   &  0.90& 1.4&   1.6& \nodata & \nodata &  16.8&  8.9 & $-0.10$\\
\enddata

\tablenotetext{a}{Clumps are identified in Fig. 4. IR clump positions 
are listed in \citet{elmegreen06}}
\tablenotetext{b} { (8 $\mu$m/6 cm) is the flux density ratio 
$S_\nu$(8 $\mu$m)/$S_\nu$(6 cm).}
\tablenotetext{c}{S(H$\alpha$) in units of 
$10^{-14}$ erg cm$^{-2}$ s$^{-1}$}
\tablenotetext{d} {These are upper limits obtained by assuming 
the radio continuum emission has no nonthermal component.}
\tablenotetext{e} {(8 $\mu$m/$UVM2$) is the flux density ratio
$S_\nu$(8 $\mu$m)/$S_\nu (UVM2)$.}
\tablenotetext{f}{in H I massive cloud N6 on the {\it NE radio ridge} 
of NGC 2207}
\tablenotetext{g}{{\it Feature i}}
\tablenotetext{h}{contains SN 1999ec}
\end{deluxetable}

\begin{deluxetable} {ccccccccc}
\tabletypesize{\footnotesize}
%\tablenum{4}
\tablewidth{0pt}
\tablecaption{H II Regions in M81\label{m81}}
\tablehead{
\colhead{H II Region\tablenotemark{a}} 
&\colhead{$S_\nu$(6 cm)}
&\colhead{(8 $\mu$m/6 cm)\tablenotemark{b}}  
&\colhead{(24 $\mu$m/6 cm)\tablenotemark{c}}
& \colhead{(8 $\mu$m/$NUV)$\tablenotemark{d}}
&\colhead{$A_v$(PG)}
&\colhead{$A_v$(K)}
\\
&\colhead{(mJy)}&&&& \colhead{(mag)} & \colhead{(mag)}
}
\startdata
K181 & $1.10 \pm 0.07$ & $18 \pm 1$ & $52 \pm 3$ & 12 & 0.70 & $0.5 \pm 0.2$\\
K178 & $1.01 \pm 0.07$ & $15 \pm 1$ & $29 \pm 2$ & 31 & 1.33 & $1.2 \pm 0.2$\\
K123 & $0.87 \pm 0.12$ & $13 \pm 2$ & $25 \pm 3$ & 21 & 0.87 & $1.0 \pm 0.3$\\
K152 & $0.80 \pm 0.14$ & $15 \pm 3$ & $34 \pm 6$ & 15 & 0.42 & $0.6 \pm 0.3$\\
K125 & $0.74 \pm 0.14$ & $16 \pm 3$ & $30 \pm 6$ & 15 & 0.43 & $0.5 \pm 0.3$\\
K159 & $0.69 \pm 0.13$ & $19 \pm 4$ & $42 \pm 8$ & 17 & 1.01 & $1.0 \pm 0.3$\\
K138 & $0.60 \pm 0.09$ & $19 \pm 3$ & $48 \pm 7$ & 12 & 0.67 & $0.7 \pm 0.3$\\
K172 & $0.52 \pm 0.07$ & $20 \pm 3$ & $35 \pm 5$ & 9.0& 0.73& $-0.1 \pm 0.3$\\
K156 & $0.50 \pm 0.10$ & $21 \pm 4$ & $25 \pm 5$ & 48 & 1.91 & $2.1 \pm 0.4$\\
K187 & $0.45 \pm 0.08$ & $29 \pm 5$ & $38 \pm 7$ & 29 & 0.36 & $0.2 \pm 0.3$\\
K102 & $0.27 \pm 0.05$ & $28 \pm 5$ & $41 \pm 8$ & 23 & 0.88 & $0.9 \pm 0.3$\\
 & mean\tablenotemark{e}
& $19 \pm 5$ & $36 \pm 9$ & $21 \pm 11$ & $0.85 \pm 0.46$ & $0.8 \pm 0.6$\\
\enddata

\tablenotetext{a}{The radio continuum data and $A_v$(K) from 
\citet{kaufman87} 
and the {\it Spitzer} data, {\it GALEX} $NUV$ data, and 
$A_v$(PG) from \citet{perez06}}
\tablenotetext{b} { (8 $\mu$m/6 cm) is the flux density ratio 
$S_\nu$(8 $\mu$m)/$S_\nu$(6 cm).}
\tablenotetext{c} { (24 $\mu$m/6 cm) is the flux density ratio 
$S_\nu$(24 $\mu$m)/$S_\nu$(6 cm).}
\tablenotetext{d} { (8 $\mu$m/$NUV$) is the flux density ratio 
$S_\nu$(8 $\mu$m)/$S_\nu(NUV)$.}
\tablenotetext{e}{The uncertainty listed with the mean is the standard
deviation $\sigma$ of the sample, not the measurement uncertainty or the
standard deviation of the mean.}
\end{deluxetable}

\begin{deluxetable}{llccccc}
\tabletypesize{\footnotesize}
%\tablenum{5}
\tablewidth{0pt}
\tablecaption{Comparison of $q_{{\rm IR}}$ Values \label{q}}
\tablehead{
\colhead{Sample\tablenotemark{a}} 
&\colhead{$q_{FIR}$}
&\colhead{$q_{70}$}
&\colhead{$q_{24}$}
&\colhead{$q_8$}
&\colhead{(24 \micron/8 \micron)\tablenotemark{b}}
}
\startdata
NGC 2207/IC 2163    &   1.81  &   1.79  &   0.71   &   0.49  & 1.6 \\
Spitzer First-Look: \\
$\;$ \citet{appleton04} & \nodata & $2.15 \pm 0.16$ & $0.94 \pm 0.23$ & 
\nodata & \nodata \\
$\;$ \citet{wu05} & \nodata & \nodata & $1.07 \pm 0.17$ & $0.91 \pm 0.13$ 
 & \nodata \\
SINGS\tablenotemark{c} &  \nodata & $2.39 \pm 0.28$ & $1.31 \pm 0.31$
& $1.12 \pm 0.26$ & $1.7 \pm 1.6$\\
SINGS\tablenotemark{d} &  \nodata & $2.40 \pm 0.29$ & $1.33 \pm 0.31$
& $1.15 \pm 0.24$ & $1.5 \pm 1.1$ \\ 
IRAS \citep{condon92} &  $2.3 \pm 0.2$ & \nodata & \nodata & \nodata
& \nodata \\
M81 H II regions & \nodata & \nodata & $1.50 \pm 0.10$ & $1.22 \pm 0.12$ 
& $1.9 \pm 0.5$\\
\enddata

\tablenotetext{a} {The Spitzer First-Look, SINGS, and IRAS  samples refer to
the integrated emission from the entire galaxy. The M81 H II region sample
refers to the H II regions in Table\,\ref{m81}. The uncertainties listed are
the standard deviations $\sigma$ of the samples.}
\tablenotetext{b} { (24 \micron/8 \micron) is the flux density
ratio $S_\nu$(24 \micron)/$S_\nu$(8 \micron).}
\tablenotetext{c} {excluding galaxies with poor quality data 
or low metallicity}
\tablenotetext{d} {excluding galaxies with poor quality data or
low metallicity and also omitting E and S0 galaxies}
\end{deluxetable}  

\clearpage

\begin{deluxetable} {lcccccc}
\tabletypesize{\footnotesize}
%\tablenum{6}
\tablewidth{0pt}
\tablecaption{Integrated \sixcm\ Radio and Infrared Flux Densities
\label{totalflux}}
\tablehead{
\colhead{Region}
&\colhead{$S_\nu$(6 cm)}
&\colhead{$S_\nu$(8 \micron)}
&\colhead{$S_\nu$(24 \micron)}
&\colhead{(8 \micron/6 cm)\tablenotemark{a}}
&\colhead{(24 \micron/6 cm)\tablenotemark{b}}
& \colhead{(24 \micron/8 \micron)\tablenotemark{c}}\\
& \colhead{(mJy)} &\colhead{(mJy)} & \colhead{(mJy)} & &\\
}
\startdata
NE radio ridge box & $13.2 \pm 0.2$ & 121 & 
 112\tablenotemark{d} & 9.1 &  8.5\tablenotemark{d} 
& 0.9\tablenotemark{d}  \\
eyelid box         & $21.7 \pm 0.3$ & 409 & 506  & 19 & 23 &  1.2\\
combined galaxies  & $132 \pm 1$ & $1.22 \times 10^3$ & $2.00 \times 10^3$ &
9.3 & 15 & 1.6\\
M81 H II regions  & \nodata & \nodata & \nodata & $19 \pm 5$ & 
$36 \pm 9$ & 1.9\\
\enddata

\tablenotetext{a} { (8 $\mu$m/6 cm) is the flux density ratio 
$S_\nu$(8 $\mu$m)/$S_\nu$(6 cm).}
\tablenotetext{b} { (24 $\mu$m/6 cm) is the flux density ratio 
$S_\nu$(24 $\mu$m)/$S_\nu$(6 cm).}
\tablenotetext{c} { (24 \micron/8 \micron) is the flux density
ratio $S_\nu$(24 \micron)/$S_\nu$(8 \micron).}
\tablenotetext{d} {uncertain because the 14\arcsec width of the 
{\it NE radio ridge box} is only
$2.3 \times$ the FWHM of the 24 \micron\ PSF.}
\end{deluxetable}

\begin{deluxetable}{lcccccccc}
\tabletypesize{\footnotesize}
\tablecaption{X-ray spectral analysis of X-ray sources
 displayed in
Figure\,\ref{n2207_xray_image}\tablenotemark{a} 
\label{source_list}}
\tablewidth{0pt}
\tablehead{
\colhead{Source} & \colhead{$\rm RA_{2000}$\tablenotemark{b}} &
\colhead{$\rm Dec_{2000}$\tablenotemark{c}} & 
\colhead{$N_{\rm H,intr}$\tablenotemark{d}} &
\colhead{\ax\tablenotemark{e}} &
\colhead{$kT$\tablenotemark{f}} &
\colhead{$\chi^2/{\rm dof}$} &
\colhead{$F_{\rm 0.3-10.0}$\tablenotemark{g}} &
\colhead{$N$(HI)\tablenotemark{h}}
} 
\startdata
 X1 & 06 16 17.94 & --21 22 04.5 & 0.12$^{+0.21}_{-0.12}$ & 
1.06$^{+0.47}_{-0.59}$ & \nodata & 20/19 & 4.70 & $0.32 \pm 0.05$ \\
 X2 & 06 16 15.86 & --21 22 08.5 & 0.00 & 2.17$^{+1.62}_{-0.56}$ & 
\nodata & 10/6 & 5.40 
& $0.32 \pm 0.08$ \\
 X3 & 06 16 18.83 & --21 22 30.5 &  0.03$^{+0.04}_{-0.04}$ & \nodata & 
2.12$^{+0.68}_{-0.68}$ & 16/16 & 1.39 & 
$0.28 \pm 0.05$ \\
 X4 & 06 16 20.34 & --21 22 16.5 & 0.12$^{+0.21}_{-0.12}$ 
& 0.93$^{+0.49}_{-0.50}$ & \nodata & 16/17 & 4.03 & $0.12 \pm 0.03$  \\
 X5\tablenotemark{i} & 06 16 21.89 & --21 22 25.5 & \nodata  
& \nodata  & \nodata & \nodata & \nodata & $0.14 \pm 0.04$ \\
 X6 & 06 16 15.84 & --21 22 34.6 & 0.15$^{+0.15}_{-0.13}$ 
& 0.97$^{+0.46}_{-0.30}$ & \nodata & 11/10 & 8.82 & $0.23 \pm 0.11$ \\
 X7 & 06 16 17.29 & --21 22 53.2 & 0.27$^{+0.42}_{-0.24}$ &
  0.86$^{+0.58}_{-0.41}$ & \nodata & 20/13 & 10.4 & $0.27 \pm 0.10$ \\
 X8 & 06 16 23.47 & --21 22 18.5 & 0.03$^{+0.12}_{-0.03}$ 
& 1.13$^{+0.51}_{-0.13}$ & \nodata & 20/17 & 4.30 & $0.27 \pm 0.10$ \\
 X9 & 06 16 24.95 & --21 22 28.9 & 0.04$^{+0.15}_{-0.04}$ 
& 1.05$^{+0.63}_{-0.43}$ & \nodata & 15/17 & 3.44 & $0.41 \pm 0.06$ \\
X10 & 06 16 26.50 & --21 22 13.4 & 0.35$^{+0.27}_{-0.15}$ 
& 1.03$^{+0.40}_{-0.27}$ & \nodata & 18/24 & 9.80 & $0.44 \pm 0.08$  \\
 RR\tablenotemark{j} & 06 16 25.13 & --21 22 23.4 & 0.09$^{+0.19}_{-0.09}$ 
& 1.14$^{+0.83}_{-0.55}$ & \nodata & 10/17 & 3.85 & \nodata
\enddata

\tablenotetext{a}{For all sources the EPIC pn data were used except for X5 
(NGC 2207 nucleus) for which the pn plus MOS data were fitted simultaneously in
XSPEC (as shown in Figure\,\ref{n2207_nucleus_xray_spec}),
and sources X2 and X7 for which only the MOS data were used because both
sources were in a gap between two CCDs in the pn. All fits include 
absorption by
neutral gas in our Galaxy with an absorption column density $N_{\rm
H,gal}=1.13\times10^{21}$ cm$^{-2}$ \citep{dic90}.
}
\tablenotetext{b}{RA in h m s}
\tablenotetext{c}{Dec in $^{\circ}$, $^{'}$, and $^{''}$}
\tablenotetext{d}{Intrinsic absorption column density at the location of NGC
2207/IC 2163 (z=0.00941) in units of $10^{22}$ cm$^{-2}$.}
\tablenotetext{e}{Energy spectral index \ax for a single, absorbed, power-law
model}
\tablenotetext{f}{$kT$ in units of keV for Raymond-Smith plasma}
\tablenotetext{g}{Flux in 0.3-10.0 keV band corrected for Galactic and 
intrinsic absorption, in units of $10^{-14}$ ergs s$^{-1}$ cm$^{-2}$}
\tablenotetext{h} {$N$(HI) of the galaxy pair in units of 
$10^{22}$ atom cm$^{-2}$ from the 21-cm line data of 
\citet{elmegreen95a} }
\tablenotetext{i}{X5 = nucleus of NGC 2207. The simple absorbed power
law model does not
represent the data. More complicated spectral models are listed in
Table\,\ref{x5_xspec}.
}
\tablenotetext{j}{RR is the {\it NE radio ridge}, not a discrete source}
\end{deluxetable}
\clearpage

\begin{deluxetable}{lcccccc}
\tabletypesize{\footnotesize}
\tablecaption{X-ray Spectral fits to the NGC 2207 
nucleus (X5)\tablenotemark{a} 
\label{x5_xspec}}
\tablewidth{0pt}
\tablehead{
\colhead{XSPEC Model} & 
\colhead{\ax\tablenotemark{b}} &
\colhead{$N_{\rm H,pc}$\tablenotemark{c}} &
\colhead{$f_{\rm pc}$} &
\colhead{kT\tablenotemark{d}} &
\colhead{$\chi^2/{\rm dof}$} &
\colhead{$F_{\rm 0.3-10.0 keV}$\tablenotemark{e}} 
} 
\startdata
zpcfabs * powl & 1.00 (fixed) & 26.5$^{+5.5}_{-6.0}$ & 0.95$^{+0.02}_{-0.01}$ &
\nodata & 83/65 & 20.6 \\
bb + powl & $-1.56^{+0.14}_{-0.16}$ & \nodata & \nodata & 163$^{+29}_{-19}$ & 
74/62 & 13.5\\ 
\enddata

\tablenotetext{a}{All fits include 
absorption by
neutral gas in our Galaxy with an absorption column density $N_{\rm
H,gal}=1.13\times10^{21}$ cm$^{-2}$ \citep{dic90}.
}
\tablenotetext{b}{Energy spectral index \ax}
\tablenotetext{c}{Intrinsic absorption column density at the location of NGC
2207/IC 2163 (z=0.00941) in units of $10^{22}$ cm$^{-2}$.}
\tablenotetext{d}{Blackbody temperature kT in units of eV} 
\tablenotetext{e}{0.3-10.0 keV flux corrected for Galactic and intrinsic
absorption, in units of $10^{-14}$ ergs s$^{-1}$ cm$^{-2}$}
\end{deluxetable}

\begin{deluxetable} {llllll}
\tabletypesize{\footnotesize}
\tablewidth{0pt}
\tablecaption{Gaussian Fits to $\lambda 6$ cm Emission from Feature i
\label{featuri_model}}
\tablehead{
\colhead{Beam}
&\colhead{Core}
&&&\colhead{Envelope}&\\
\colhead{HPBW}
&\colhead{$S_\nu$(6 cm)}
&\colhead{FWHM, PA}
&&\colhead{$S_\nu$(6 cm)}
&\colhead{FWHM, PA}\\
&\colhead{(mJy)} &&&\colhead{(mJy)}&\\
}
\startdata
$2.50'' \times 2.50''$   & $2.90 \pm 0.03$ & $1.4'' \times 1.0''$, 39\arcdeg\
  & & $1.69 \pm 0.05$ & $5.6'' \times 1.9''$, 142\arcdeg \\
$2.48'' \times 1.30''$  & $2.66 \pm 0.02$ & $1.2'' \times 0.8''$, 43\arcdeg\
   && $2.00 \pm 0.07$ & $5.5'' \times 2.6''$, 145\arcdeg\\
\enddata
\end{deluxetable}

\end{document}